\documentclass[a4paper,fleqn,usenatbib]{mnras}

\title[Magnetic buoyancy in simulated galactic discs]{Magnetic buoyancy in simulated galactic discs with a realistic circum galactic medium}
\author[U. P. Steinwandel et al.]{U. P. Steinwandel$^{1,2,3}$\thanks{E-mail: usteinw@usm.lmu.de}, M. C. Beck$^{3}$, A. Arth$^{1,4}$, K. Dolag$^{1,2}$, B. P. Moster$^{1,2}$
\newauthor and P. Nielaba$^{3}$ \\
  $^{1}$Universit\"{a}ts-Sternwarte M\"{u}nchen, Fakult\"{a}t f\"{u}r Physik, LMU Munich, Scheinerstr. 1, D-81679, Germany \\
  $^{2}$Max Planck Institute for Astrophysics, Karl-Schwarzschild-Str. 1, D-85741 Garching, Germany \\
  $^{3}$Department of Physics, University of Konstanz, Universit\"{a}tsstr. 10, D-78457 Konstanz, Germany \\
  $^{4}$Max Planck Institute for Extraterrestrial Physics, Giessenbachstr. 1, D-85748 Garching, Germany} 
\usepackage{newtxtext,newtxmath}
\usepackage[T1]{fontenc}
\usepackage{ae,aecompl}
\usepackage{pdfpages}
\usepackage{graphicx}
\usepackage{aas_macros}

\usepackage{amsmath}	
\usepackage{amssymb}	

\usepackage{color}

\usepackage{changes}
\definechangesauthor[color=orange]{MB}
\definechangesauthor[color=red]{AA}
\usepackage{todonotes}

\hypersetup{draft}

\begin{document}

\date{Accepted XXX. Received YYY; in original form ZZZ}
\pagerange{\pageref{firstpage}--\pageref{lastpage}} \pubyear{2018}
\maketitle
\label{firstpage}


\begin{abstract}
We present simulations of isolated disc galaxies in a realistic environment performed with the Tree-SPMHD-Code \textsc{Gadget-3}.
Our simulations include a spherical circum-galactic medium (CGM) surrounding the galactic disc, motivated by observations and the results of cosmological simulations.
We present three galactic models with different halo masses between $10^{10} M_\odot$ and $10^{12} M_\odot$, and for each we use two different approaches to seed the magnetic field, as well as a control simulation without a magnetic field. 
We find that the amplification of the magnetic field in the centre of the disc leads to a biconical magnetic outflow of gas that magnetizes the CGM.
This biconical magnetic outflow reduces the star formation rate (SFR) of the galaxy by roughly $40 \%$ compared to the simulations without magnetic fields.
As the key aspect of our simulations, we find that small scale turbulent motion of the gas in the disc leads to the amplification of the magnetic field up to tens of $\mu G$, as long as the magnetic field strength is low.
For stronger magnetic fields turbulent motion does not lead to significant amplification but is replaced by an $\alpha$ - $\omega$ dynamo.
The occurance of a small scale turbulent dynamo becomes apparent through the magnetic power spectrum and analysis of the field lines' curvature.
In accordance with recent observations we find an anti-correlation between the spiral structure in the gas density and in the magnetic field due to a diffusion term added to the induction equation.
\end{abstract}




\section{Introduction}
Magnetic fields are a fundamental aspect in astrophysics and cosmology. They are essential for describing many processes in theoretical astrophysics properly. The relevance of magnetic fields ranges from the small scales in star formation and in the ISM over galactic scales up to galaxy-clusters and the large scale structure of the Universe. While there is a large amount of observational data on magnetic fields, especially in the area of galaxy formation and evolution \citep[i.e.][]{Hummel1986, Chyzy2003, Chyzy2007, Beck2007} the situation is different for numerical studies that investigate the behaviour of magnetic fields in detail
\citep{Kotarba2011, Pakmor2013, Mbeck2016, Rieder2016, Pakmor2017}.
In the case of galaxies the magnetic field becomes important for several reasons. It acts as an additional pressure component, and thus is needed as a correction of the equations of hydrodynamics, resulting in the well known equations of magnetohydrodynamics. Moreover, the magnetic pressure in galaxies can reach the same order of magnitude as the turbulent pressure of the ISM, and thus it can
completely dominate the thermal pressure of the disc. Hence, it should be taken into account in simulations of galaxy formation and evolution.  
However, this is often not the case and pure hydrodynamical simulations are used to study the formation and evolution of galactic discs. On the other hand magnetic fields can be very important for star formation and the regularization of cosmic rays and should not be excluded when these processes are taken into account.

Still, the origin of magnetic fields in the Universe is unclear. It is possible to generate magnetic seed fields below $10^{-20}$ G by either battery processes in the early Universe \citep[e.g.][]{Biermann1950, Mishustin1972, Zeldovich1983} or the phase transitions that appear in the standard model, shortly after the Big Bang \citep[e.g.][]{Hogan1983, Ruzmaikin1988a, Ruzmaikin1988b, Widrow2002}. 
Those initial magnetic fields can then be amplified by various dynamo processes, namely the $\alpha$-$\omega$-dynamo \citep{Ruzmaikin1979}, the cosmic ray driven dynamo \citep{Lesch2003, Hanasz2009} or the small scale turbulent dynamo \citep{Kazantsev1968, Kraichnan1968}. While, the first and the second dynamo process operate on $10^{8}$ yr-scales, the third process has typical timescales in the Myr-regime \citep{Biermann1951}. The magnetic energy is exponentially amplified on the small scales and random motion regulates it on the largest turbulent scales \citep{Zeldovich1983, Kulsrud1992, Kulsrud1997, Malyshkin2002, Schekochihin2002, Schekochihin2004, Schleicher2010}.

Observationally, there are several methods to measure the magnetic field strengths in galaxies. \citet{Brown2007} investigated the magnetic field of the inner Milky Way by using rotation measurements of $148$ objects behind the galactic disc. Further, magnetic fields of nearby galaxies can be determined using  their radio synchrotron emission. In this case the unpolarized component of the synchrotron emission is important because it is needed to explain galactic dynamics and magnetic driven outflows \citep{Beck2007}. Radio synchrotron emission is also used to calculate the magnetic field strengths in nearby galaxies which leads to values between $20$ and $30$ $\mu$G in the spiral arms and to $50$ to $100$ $\mu$G in the galactic centre, as described in \citet{Beck2007} or \citet{Beck2009}. \citet{Robishaw2008} presented measurements of magnetic fields due to Zeeman-splitting emission in OH megamasers of five ultra luminous infrared galaxies leading to magnetic field strengths along the line of sight between $0.5$ and $18$ mG. Beyond the observations of the magnetic fields in galactic discs there are also those of their CGM. \citet{Carretti2013} present measurements of magnetized outflows towards the CGM of the Milky Way in two giant lobes located in the north and south of the galactic centre with a magnetic field strength of around $15$ $\mu$G, known as the 'Fermi-bubbles'.

Simulations of galactic magnetic fields are often done with hydrodynamical grid codes.
\citet{Wang2009} investigated the magnetic field in isolated disc galaxies without star formation using the grid code \textsc{enzo} \citep{Bryan1997, Oshea2004, Bryan2014}. \citet{Dubois2010} studied the magnetic field of dwarf galaxies with a closer look on winds driven by stars using the grid code \textsc{ramses} \citep{Teyssier2002}. \citet{Pakmor2013} and \citet{Rieder2016} present studies of magnetic fields for isolated galaxy formation by collapsing a giant gaseous halo in a dark matter potential using the moving mesh code \textsc{arepo} \citep{Springel2010} and
the grid code \textsc{ramses}, respectively. While \citet{Pakmor2013} investigate general properties of the magnetic field, \citet{Rieder2016} point out the importance of supernova feedback on the structure of the ISM and its magnetic field.
Another detailed study of magnetic fields in isolated discs is presented in \citet{Butsky2017} where they find a small scale turbulent dynamo of the magnetic field. The same behaviour can be found in \citet{Rieder2017a} for an isolated disc galaxy as well as in \citet{Pakmor2017} and 
\citet{Rieder2017b} for cosmological zoom-in simulations.  The evolution of magnetic fields has been extensively studied with particle-based methods for magnetohydrodynamics as well.
\citet{Kotarba2009} investigate the magnetic field in an isolated disc galaxy using the SPH-code \textsc{vine} \citep{Wetzstein2009}. In \citet{Kotarba2010} \textsc{vine} is used for studies of the magnetic field in the Antennae-galaxies. In \citet{Kotarba2011}, \citet{Ageng2012a}, and in \citet{Ageng2012b}, \textsc{gadget-3} is used to study the magnetic field in other galaxy mergers. \citet{Mbeck2016} is investigating
the magnetic field structure of the Milky Way in more detail, by calculating its synchrotron-emission. Many of these simulations study the evolution of galaxies in isolation by sampling a dark matter halo, a stellar bulge, and a stellar and gaseous disc to create the initial conditions. The hot CGM is typically neglected to save computational effort and as it is mainly irrelevant for galactic dynamics.
However, cosmological simulations show that galaxies are constantly accreting gas from their hot haloes, such that this component is essential to model realistic galactic systems. Moreover, a hot gaseous halo around the Milky Way can now be detected observationally \citep{miller13}. These observations indicate that the density profile of the CGM can be described with a $\beta$-power law \citep{cavaliere78}, which is common in studies of globular clusters \citep{Plummer1911} and cosmological simulations of galaxy-clusters  \citep[e.g.][]{donnert14}. Consequently, it may be critical to include the hot gaseous component in simulations of isolated galaxies.
In SPMHD simulations the presence of the CGM has a further advantage. In the SPMHD formalism the magnetic field is a property of the gas particles. Therefore, a carrier for the magnetic field is needed which gives further justification for the presence of the CGM in magnetohydrodynamic simulations. 
In this work we present a set of nine high-resolution simulations that include a spherical hot CGM for each galaxy. The paper is structured as follows. We give a short summary of our simulation method in section \ref{sec:simulation_method}, where we point out recent improvements of our numerical methods and the physical models that are considered. In section \ref{sec:initial_conditions} we present the methods that we use to build our numerical model for an isolated disc galaxy that includes an observationally constrained CGM.
We then examine the general properties of each galactic system and investigate the interaction between the galactic disc and the CGM in section \ref{sec:results}. Our conclusions are presented in section \ref{sec:conclusions}.


\section{Simulation Method} \label{sec:simulation_method}

The simulations we present in this paper are performed using the Tree-SPMHD-Code \textsc{gadget}-3, the developers version of the public available \textsc{gadget}-2 code \citep{springel05}. We use a modern implementation of SPH, as presented in \citet{Beck2016}. This SPH formulation includes various improvements like a higher order SPH-kernel described by \citet{dehnen12}, a timestep limiter \citep{Dalla2012}, time-dependent artificial viscosity and a new model for time-dependent artificial conduction. Our version of \textsc{gadget-3} further includes magnetic fields and magnetic dissipation as presented in \citet{dolag09}. We also include star formation, cooling, supernova-feedback and metals following \citet{springel03}. Further, one of our models for the magnetic field couples the seed rate of the magnetic field directly to the supernova rate within the ISM, as presented in \citet{abeck13a}. In this section we summarize the adopted SPH formalism and the physical models in a very compact way.

\subsection{Kernel function and density estimate}

We use the density-entropy formulation of SPH, i.e. we smooth the density distribution such that
\begin{align}
  \rho_{i} = \sum_{j} m_{j} W_{ij}(x_{ij}, h_{i}), 
  \label{eq:sph}
\end{align}
where $h_{i}$ is the smoothing-length. The sum in equation \ref{eq:sph} is calculated over the neighbouring particles. $W_{ij}(x_{ij}, h_{i})$ is the smoothing kernel with the property 
\begin{align}
  W_{ij}(x_{ij}, h_{i}) = \frac{1}{h_{i}^3} w(q). 
\end{align} 
In our simulations we use the Wendland C4 function for $w(q)$, with 200 neighbouring particles. The function $w(q)$ is given by 
\begin{align}
  w(q) = \frac{495}{32 \pi} (1-q)^6 \left(1+ 6q + \frac{35}{3} q^2\right), 
\end{align}
for $q<1$. For $q>1$ we set $w(q)$ to zero.

\subsection{SPH and SPMHD formulation}

It is possible to derive the equations of motion (EOM) in both the hydrodynamical and the magnetohydrodynamical case from a discrete Lagrangian, via the principle of least action, as been presented in \citet{price12}. This leads to the SPH-formulation of the EOM:
\begin{align}
  \frac{\mathrm{d} \mathbf{v}_{i}}{\mathrm{d}t} = -\sum_{j} m_{j} \left[f_{i}^\mathrm{co} \frac{P_{j}}{\rho_{j}^{2}} \frac{\partial W_{ij}(h_{i})}{\partial \mathbf{r}_{i}} + f_{j}^\mathrm{co} \frac{P_{j}}{\rho_{j}^{2}} \frac{\partial W_{ij}(h_{j})}{\partial \mathbf{r}_{i}} \right].
\label{eq:EOMsph}
\end{align}   
with $f_{j}^\mathrm{co}$ given by
\begin{align}
  f_{j}^\mathrm{co} = \left[1 + \frac{h_{j}}{3 \rho_{j}} \frac{\partial \rho_{j}}{\partial h_{j}}\right]^{-1}.
\end{align}
The formulation given by equation \ref{eq:EOMsph} conserves energy, momentum and angular momentum per construction.
For the case of SPMHD, the EOM take the form
\begin{align}
  \frac{\mathrm{d}\mathbf{v}_{i}}{\mathrm{d}t} = -\sum_{j} m_{j} \left[f_{i}^\mathrm{co} \frac{P_{i} + \frac{1}{2\mu_{0}} B_{i}^{2}}{\rho_{i}^{2}} \nabla_{i} W_{ij}(h_{i}) \right. \notag \\
  \left. + f_{j}^\mathrm{co} \frac{P_{j} + \frac{1}{2\mu_{0}} B_{j}^{2}}{\rho_{j}^{2}}  \nabla_{i} W_{ij}(h_{j}) \right] \notag \\
      + \frac{1}{\mu_{0}} \sum_{j} m_{j} \left[f_{i}^\mathrm{co} \frac{\textbf{B}_{i}[\textbf{B}_{i} \cdot \nabla_{i} W_{ij}(h_{i})]}{\rho_{i}^{2}} \right. \notag \\
      \left. + f_{j}^\mathrm{co}\frac{\textbf{B}_{j}[\textbf{B}_{j} \cdot \nabla_{i} W_{ij}(h_{j})]}{\rho_{j}^{2}}\right]. 
  \label{eq:SPMHD}
\end{align}

We note, that the magnetic field influences the EOM in two ways. At first the presence of the magnetic field generates a pressure alongside the thermal pressure of the fluid, which scales like $\mathbf{B}^2$. The second term is needed to fulfil the $\nabla \cdot \textbf{B}=0$ constraint. Further, we note that for the SPMHD formulation, energy and linear momentum are conserved down to machine precision, while angular momentum is violated due to the fact that the second term in equation \ref{eq:SPMHD} is anisotropic and therefore not invariant under rotation of the system. 

\subsection{Cooling, star formation and supernova-seeding} \label{sec:models}

We briefly describe the cooling, star formation and the supernova-seeding approach that is used in a subset of our simulations. We include cooling as decribed by \citet{Katz1996}. In this framework, cooling is mainly driven due to collisional excitation of H$^{0}$ and He$^{+}$, and free-free emission (thermal Bremsstrahlung). The cooling rates are then calculated via the assumption of collisionless ionization equilibrium and an optically thin gas. We use the stochastic star formation approach, presented in \citet{springel03}, where stars are formed according to the Kennicutt-Schmidt relation \citep{Schmidt1959, Kennicutt1989}. The adopted values for the star formation model are given in Table \ref{tab:multiphase_model}. Simulations without magnetic fields are referred to as \textit{noB}. Additionally, we perform runs with two different magnetic field models. The first one is set up with a primordial magnetic field of $10^{-9}$G in x-direction in the disc and $10^{-12}$G in the CGM, and is referred to as \textit{primB}.The second model does not employ a primordial magnetic field, but follows the magnetic supernova-seeding presented in \citet{abeck13a}, and is referred to as \textit{snB}. Here, a dipole field is seeded in the ISM when a supernova explodes, such that it directly couples to the stellar feedback routines. The induction equation is modified with an additional seeding term on the right hand side:
\begin{align}
  \frac{\partial \mathbf{B}}{\partial t} = \nabla \times \left( \mathbf{v} \times \mathbf{B} \right) +\eta \Delta \mathbf{B} + \left(\frac{\partial \mathbf{B}}{\partial t}\right)_\mathrm{seed},
\end{align}
with the magnetic resistivity $\eta$ and the magnetic seeding amplitude per timestep, calculated via
\begin{align}
  \left(\frac{\partial \mathbf{B}}{\partial t}\right)_\mathrm{seed} = \sqrt{N_\mathrm{SN}^{\mathrm{eff}}} \, \frac{B_\mathrm{Inj}}{\Delta t} \, \mathbf{e}_\mathrm{B},
\end{align}
where $N_\mathrm{SN}^{\mathrm{eff}}$ is the effective number of supernovae directly given by the \citet{springel03} star formation model. The parameter $\mathbf{e}_\mathrm{B}$ is a normalization vector and $B_\mathrm{Inj}$ is the injected magnetic field amplitude given by
\begin{align}
  B_\mathrm{Inj} = \sqrt{N_\mathrm{SN}^{\mathrm{eff}}} \, B_\mathrm{SN} \, \left(\frac{r_\mathrm{SN}}{r_\mathrm{SB}}\right)^2 \left(\frac{r_\mathrm{SB}}{r_\mathrm{r_\mathrm{Inj}}}\right)^{3},
\end{align}
with $B_\mathrm{SN} = 10^{-5}$ G as the magnetic seed field strength within the supernova-radius $r_\mathrm{SN} = 5$ pc. The bubble radius $r_\mathrm{SB} = 25$ pc is the radius where the isotropic expansion of the magnetic field within the shell ends.
The bubbles are randomly placed within the injection radius $r_\mathrm{Inj}$ and mix with the surrounding medium.

  \begin{table}
    \centering
    \caption{Parameters for the multiphase model.}
    \label{tab:multiphase_model}
    \begin{tabular}{lll}
      \hline
      \multicolumn{3}{c}{Multiphase model parameters}\\\hline
      Gas consumption time-scale in [Gyr]             &$t_\mathrm{MP}$             &$2.1$ \\
      Mass fraction of massive stars                  &$\beta_\mathrm{MP}$         &$0.1$\\
      Evaporation parameter                           &$A_{0}$                	   &$1000$\\
      Effective supernova temperature in [K]          &$T_\mathrm{SN}$             &$1 \cdot 10^8$ \\
      Temperature of cold clouds in [K]               &$T_\mathrm{CC}$             &$1000$ \\
      \hline
    \end{tabular}
  \end{table}


\section{Initial conditions} \label{sec:initial_conditions}
 
Our setup consists of an isolated disc galaxy, surrounded by a spherical CGM.
To set the system up, we use the method described in \citet{hernquist93}. A more detailed documentation can be found in \citet{springel99} and \citet{springel05b}. The model for a spiral disc galaxy consists of a dark matter halo, a bulge, a stellar disc and a gaseous disc.   
We prepare initial conditions for three systems with virial masses of $10^{10}$, $10^{11}$, and $10^{12} \mathrm{M}_{\odot}$,
representing a dwarf galaxy (DW), a medium-mass galaxy (MM), and a Milky Way-like galaxy (MW), respectively. In Table \ref{tab:particle_numbers} we present the particle numbers used in our models and list the mass resolution as well as the gravitational softening lengths.

\subsection{Galactic system}

  \begin{table}
    \centering
    \caption{Number of particles, mass resolution, and gravitational softening lengths for our three galactic systems.}
    \label{tab:particle_numbers}
    \begin{tabular}{llrrr}
      \hline
      \multicolumn{5}{c}{Particle Numbers $[10^{6}]$}\\\hline
				&			& DW &  MM & MW \\
      Gas disc			&$N_\mathrm{gd}$	& $0.8$ & $1.0$ & $1.2$\\
      Gas halo			&$N_\mathrm{gh}$	& $5.0$ & $6.0$ & $7.0$\\
      Stellar disc		&$N_\mathrm{sd}$	& $3.2$ & $4.0$ & $4.8$\\
      Stellar bulge		&$N_\mathrm{b}$		& $1.3$ & $1.6$ & $2.0$\\
      Dark matter		&$N_\mathrm{dm}$	& $4.6$ & $5.7$ & $6.9$\\
      \hline

      \hline
      \multicolumn{5}{c}{Mass resolution $[M_{\odot}]$}\\\hline
                                &                               & DW & MM & MW\\
      Gas particles             &$m_\mathrm{gas}$   & 72 & 510 & 4800\\
      Star particles            &$m_\mathrm{star}$  & 72 & 510 & 4800\\
      Dark matter               &$m_\mathrm{dm}$    & 1440 & 10200 & 96000\\
      \hline
      \multicolumn{5}{c}{Gravitational softening $[pc]$}\\\hline
				&				& DW & MM & MW\\
      Gas particles		&$\epsilon_\mathrm{gas}$	& 5 & 10 & 20\\
      Star particles		&$\epsilon_\mathrm{star}$	& 5 & 10 & 20\\
      Dark matter		&$\epsilon_\mathrm{dm}$		& 40 & 20 & 10\\
      \hline
    \end{tabular}
  \end{table}
  
The dark matter halo is modelled using the spherical \citet{hernquist93} density profile:
\begin{equation}
	\rho_\mathrm{dm}(r) = \frac{M_\mathrm{dm}}{2 \pi} \frac{a}{r(r + a)^3}, 
\end{equation}
where $M_\mathrm{dm}$ is the total mass of the dark matter halo, and $a$ is a scale parameter. For the corresponding \citet*{navarro97} profile with an equal inner density profile and a scale length of $r_\mathrm{s}$, the parameter $a$ can be related to the halo concentration $c$ by
\begin{align}
a = r_\mathrm{s} \sqrt{\,2 \,[\ln(1+c)-c/(1+c)]} \, .
\end{align}
The density profile of the stellar bulge follows the Hernquist-profile:
\begin{equation}
        \rho_\mathrm{b}(r) = \frac{M_\mathrm{b}}{2 \pi} \frac{l_\mathrm{b}}{r(r + l_\mathrm{b})^3}, 
\end{equation}
where $l_\mathrm{b}$ is the scale length of the bulge.
Its mass is given by $M_\mathrm{b} = m_\mathrm{b} M_\mathrm{200}$, where $m_\mathrm{b}$ is the dimensionless bulge mass fraction. \\
The surface densities of the stellar and gaseous discs $\Sigma_{\star}$ and $\Sigma_\mathrm{gas}$ follow an exponential profile:
\begin{equation}
	\Sigma_{\star} = \frac{M_{\star}}{2 \pi l_\mathrm{d}^2} \cdot \mathrm{exp}\left(-\frac{r}{l_\mathrm{d}}\right),
\end{equation} 
\begin{equation}
        \Sigma_\mathrm{gas} = \frac{M_\mathrm{gas}}{2 \pi l_\mathrm{d}^2} \cdot \mathrm{exp}\left(-\frac{r}{l_\mathrm{d}}\right),
\end{equation}
where $l_\mathrm{d}$ is the scale length of the disc. The mass of the disc is given by $M_\mathrm{d} = (M_{\star} + M_\mathrm{gas}) = m_\mathrm{d} M_\mathrm{200}$, 
with the dimensionless disc mass fraction $m_\mathrm{d}$. We take a fraction $f$ of the disc mass to compose the gaseous disc.
The total mass of the dark matter halo is then given by $M_\mathrm{dm} = M_\mathrm{200} - \left(m_\mathrm{b} + m_\mathrm{d} \right) \cdot M_\mathrm{200}$. 
Finally, the angular momentum of the system is determined by the spin parameter $\lambda$, as described by \citet{mo98}. 
The parameters we used for each model are listed in Table \ref{tab:disc_parameters}.

  \begin{table}
    \centering
    \caption{Adopted parameters for our three galactic systems.}
    \label{tab:disc_parameters}
    \begin{tabular}{lcccc}
      \hline
      \multicolumn{5}{c}{Disc parameters}\\\hline
                                           &                            & DW & MM & MW \\
      Total mass [$10^{10} M_{\odot}$]  &$M_\mathrm{200}$       	& 1 & 10 & 100  \\
      Virial radius [kpc]               &$r_{200}$                   & 31 & 67 & 145 \\
      Halo concentration		   &$c$			& 8 & 10 & 12 \\
      Spin parameter		           &$\lambda$			& 0.033 & 0.033 & 0.033 \\
      Disc mass fraction		   &$m_\mathrm{d}$		& 0.041 & 0.041 & 0.041 \\
      Bulge mass fraction		   &$m_\mathrm{b}$		& 0.013 & 0.013 & 0.013 \\
      Disc spin fraction		   &$j_\mathrm{d}$		& 0.041 & 0.041 & 0.041 \\
      Gas fraction			   &$f$				& 0.2 & 0.2 & 0.2 \\
      Disc scale length [kpc]           &$l_\mathrm{d}$         	& 0.8 & 1.5 & 2.1 \\ 
      Disc height		   &$z_0$			& 0.2 $l_\mathrm{d}$ & 0.2 $l_\mathrm{d}$ & 0.2 $l_\mathrm{d}$ \\
      Bulge size			   &$l_\mathrm{b}$		& 0.2 $l_\mathrm{d}$  & 0.2 $l_\mathrm{d}$ & 0.2 $l_\mathrm{d}$ \\
      \hline
    \end{tabular}
  \end{table}

\subsection{Circum galactic medium}
Our implementation of the CGM follows that of \citet{Moster2010} and \citet{donnert14}, but samples the particles from a glass distribution rather than from a random distribution.
We assume a radial symmetric density distribution for the gas medium surrounding the galaxy. For this we use the radial density profile of the beta-model \citep{cavaliere78}, that has also been found by observations \citep[i.e.][]{croston08, miller13}. The density distribution in the beta-model takes the form
  \begin{equation}
   	\label{eq:density_beta}
	\rho_\mathrm{gh} = \rho_0 \left(1 + \frac{r^2}{r^2_c} \right)^{-\frac 32 \beta} \;.
  \end{equation} 
We follow \citet{mastropietro08} and set $\beta = 2/3$, which is underpinned by recent observations by \citet{miller13}, who find a value 
for $\beta$ close to this value. We choose the central gas density $\rho_0$ to be between $10^{-27}$ and $5\cdot 10^{-26} \mathrm{g}/\mathrm{cm}^3$ , motivated by the electron density found through cosmological simulations \citep{dolag15}, as well as the observations of Milky Way's hot gaseous halo \citep{miller13}. The core radius $r_c$ has been chosen between $0.22$ and $0.33 \mathrm{kpc}$, which is again motivated by observations \citep{miller13}.
The value of $r_\mathrm{c}$ conforms well with $r_\mathrm{s} / 40$ where $r_\mathrm{s}$ is the scale length of the corresponding NFW-halo. 
The value of $\beta = 2/3$ allows us to calculate the mass distribution of the gaseous halo analytically. By integrating the density profile set by equation \ref{eq:density_beta}, the hot gas mass within a radius $r$ is given by
 \begin{equation}
	\label{eq:mass_distribution_cummulative}
	M_\mathrm{gh}(<r) = 4 \pi r_c^3 \rho_0 \left[\frac{r}{r_\mathrm{c}} - \arctan\left(\frac{r}{r_\mathrm{c}}\right)\right].
\end{equation}

 \begin{table}
    \centering
    \caption{Parameters for the gaseous halo}
    \label{tab:gas_halo_parameters}
    \begin{tabular}{ccccc}
      \hline
      \multicolumn{5}{c}{General parameters}\\\hline
                                                     &                                & DW & MM & MW \\
      Total Mass [$10^{10} M_{\odot}$]          &$M_\mathrm{gh}$            & $0.05$ & $0.5$ & $5.0$ \\ 
      Virial temperature [K]              &$T_\mathrm{vir}$                & $10^{4}$ & $10^{5}$ & $10^{6}$ \\
     \hline
      \multicolumn{5}{c}{Settings for the $\beta$-model}\\\hline
      Exponent                               &$\beta$         	               & & $2/3$ &                 \\
      Density in [g/cm$^{3}$]        &$\rho_{0}$                       &  & $5 \cdot 10^{-26}$ &  \\  
      Core radius                   &$r_\mathrm{c}$                   & $0.22 \cdot r_\mathrm{s}$ &  $0.25 \cdot r_\mathrm{s}$ &  $0.33 \cdot r_\mathrm{s}$                                   \\
    \hline
    \end{tabular}
  \end{table}

As we require the density distribution to be as close as possible to equilibrium, we sample the particle positions of the CGM with a normalized glass distribution.
The glass is constructed with the Wendland C4 kernel to be consistent with the disc galaxy simulations. We introduce the variable $q$ as
\begin{equation}
        q = \frac{M_\mathrm{gh}(<r_\mathrm{new})}{M_\mathrm{gh}},
\end{equation}

which corresponds to the hot gas mass within a radius $r_\mathrm{new}$, normalized by the total gas mass of the CGM.
To sample the particle distribution, we thus need to solve the equation
\begin{equation}
	\frac{4 \pi r_c^3 \rho_0}{M_\mathrm{gh}} \left[\frac{r}{r_\mathrm{c}} - \arctan\left(\frac{r}{r_\mathrm{c}}\right)\right] - q = 0.
\label{eq:newton_raphson}
\end{equation}

We take a normalized, equally distributed glass distribution and transform its components to spherical coordinates $r'$, $\theta$ and $\phi$.
As we can see from equation \ref{eq:mass_distribution_cummulative}, the mass distribution of the $\beta$-profile is only a perturbation in the radial coordinate. We determine the value of $q$ that corresponds to $r'$ by solving equation \ref{eq:newton_raphson} with the Newton-Raphson-Method.
 
This leads to a new radial component with a smaller value than $r'$, in agreement with the density distribution given by equation \ref{eq:density_beta}.
As this procedure distorts the initial glass sampling, the advantages of using a glass are somewhat weakened. However, the particle noise is kept low enough to justify the procedure (a more rigorous approach will be presented in Arth et al. in prep.) Further, we observe better results compared to other initial configurations (random distribution, cubic or hcp lattice). The angular coordinates stay unchanged and we can perform the transformation from spherical coordinates back to Cartesian coordinates with the new radial coordinate. As a result, we get a particle distribution for the CGM which is as close as possible to its dynamical equilibrium. In the last step the gaseous halo is balanced in the dark matter profile of the galactic disc. The condition for hydrostatic equilibrium between the dark matter halo of the galaxy and the CGM are
\begin{equation}
	\label{eq:equilibriums_condition}
	\frac{1}{\rho_\mathrm{gh}} \frac{\mathrm{d} P_\mathrm{gh}}{\mathrm{d} r} = - \frac{G \, M_\mathrm{total}(<r)}{r^2}
\end{equation}
This allows us to calculate the temperature profile of the CGM by integrating the equilibrium condition using the ideal equation of state for
an atomic gas:
\begin{equation}
        T(r) = \frac{\mu m_\mathrm{p}}{k_\mathrm{B}} \frac{G}{\rho_\mathrm{gh}} \int_{r}^{R_\mathrm{max}} \frac{\rho_\mathrm{gh}(t)}{t^2} M_{200}(< t) \, \mathrm{d}t  \, ,
\end{equation}
which leads to
\begin{align}
  T(r) = G \frac{\mu m_\mathrm{p}}{k_\mathrm{B}} \left(1 + \frac{r^2}{r_\mathrm{c}^2}\right) [M_\mathrm{dm} F_{0}(r) + 4 \pi r_\mathrm{c}^3 \rho_{0} F_{1}(r)].
\end{align}
The temperature profile consists out of two parts. The first part comes from the dark matter halo and the second part is the influence of the CGM itself. The functions $F_{0}(r)$ and $F_{1}(r)$ are given via

\begin{equation}
\begin{aligned}
  F_{0}(r) =& \frac{r_\mathrm{c}}{a^2 + r_\mathrm{c}^2} \left[\frac{\pi}{2} (a^2 - r_\mathrm{c}^2) + r_\mathrm{c} \frac{a^2 + r_\mathrm{c}^2}{a+r} \right. \\
  & \left. - (a^2 - r_\mathrm{c}^2) \arctan \left(\frac{r}{r_\mathrm{c}}\right) - r_\mathrm{c} a \ln \left( \frac{(a+r)^2}{r^2 + r_\mathrm{c}}\right)\right],
\end{aligned}
\label{eq:F0}
\end{equation}
and 
\begin{align}
  F_{1}(r) = \frac{\pi^2}{8 r_\mathrm{c}} - \frac{\arctan^2(r/r_\mathrm{c})}{2 r_\mathrm{c}} - \frac{\arctan(r/r_\mathrm{c})}{r}.
\end{align}

From an observational point of view it is useful to calculate a virial temperature for the gaseous halo $T_\mathrm{c}=T(r_\mathrm{c})$.
It can be calculated as

\begin{equation}
\begin{aligned}
	T_\mathrm{c} =& \frac{2G \mu m_\mathrm{p}}{k_\mathrm{B}}  \left\{
M_\mathrm{dm} \frac{r_\mathrm{c}^2}{(a^2+r_\mathrm{c}^2)} \cdot \left[\frac{\pi}{4 r_\mathrm{c}^2} 
 (a^2 - r_\mathrm{c}^2) + \frac{a^2 + r_\mathrm{c}^2}{a + r_\mathrm{c}} \right. \right. \\
& \left. \left. - a \, \mathrm{ln}\left(\frac{(a+r_\mathrm{c})^2}{2 r_\mathrm{c}^2} \right)\right] + \pi^2 r_\mathrm{c}^2 \rho_{0} \left(\frac{3 \pi}{8} - 1\right)\right\} \, .
\end{aligned}
\end{equation}
In table \ref{tab:gas_halo_parameters} we summarize the values used to construct realistic gaseous haloes for the three galactic system in this study. Finally we note that the CGM is truncated at $R200$.

\subsection{Combination of the galactic disc with the CGM}

Finally, we need to combine the galactic disc with the CGM in a way that keeps the initial conditions as close as possible to equilibrium. Further, we want to avoid
an overlap between the particles of the galactic disc and the CGM. Therefore, we implement a procedure to cut out the central part of the CGM and place the disc in the resulting gap. An obvious choice would be to cut out a cylinder, with the radius of the disc and the height of the disc height. This procedure has the advantage, that it is very simple. However, this method results in a relatively large gap between the disc and the CGM. Luckily, the density profile of the galactic disc and the CGM are slightly different and we can use this for selecting the part of the CGM we want to cut out, by introducing a quality condition for the density in the overlap region of galactic disc and CGM. To do so, we bring the SPH data of the galactic disc on a grid by using the triangular shaped cloud method. Then we compare the density of each grid cell to the density of the CGM. To minimize the gap between disc and gaseous halo we remove the particles of the CGM, if their density is ten percent different to that of the grid cell they are related to. The grid we use for this purpose has a spatial resolution of $20^3=8000$ cells.


\section{Results} \label{sec:results}

We perform three types of simulations with the same initial conditions but with different approaches for the magnetic field, a primordial field, a field seeded by supernovae, and, for reference, without a magnetic field. In this section we present our results for the morphology of the disc, the SFR, the growth rate of the magnetic field, the magnetic field structure and the interaction between the galactic disc and the CGM. We mainly focus on the Milky Way-like galaxies MW, MW-primB and MW-snB.

\subsection{Morphology of the galactic disc and the magnetic field \label{sec:morphology}}

\begin{figure*}
        \centering 
        \includegraphics[scale=0.55]{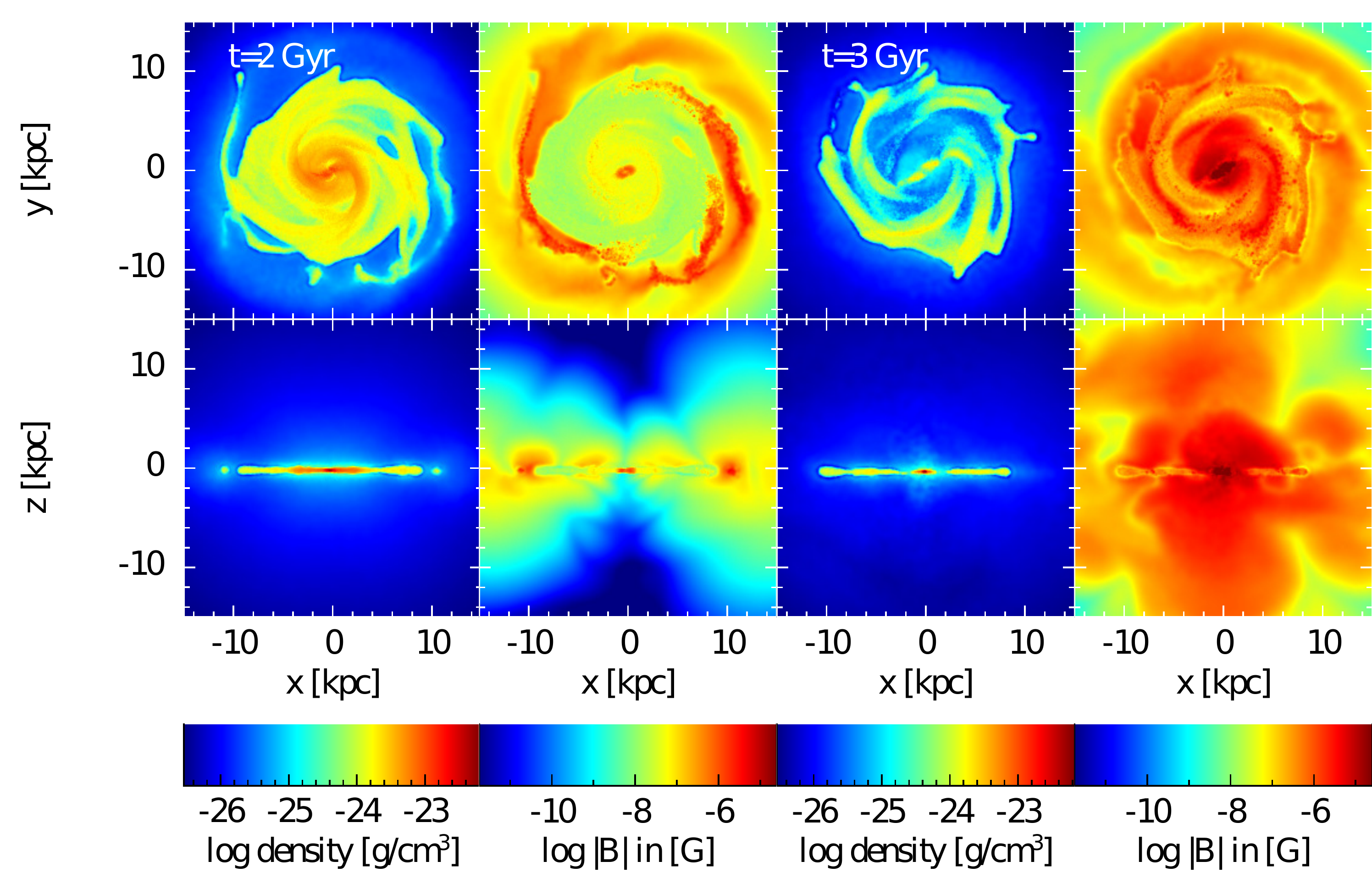}
        \caption{Slice through the gas density and the magnetic field strength for the simulation \textit{MW-snB} for the face-on and edge-on view. The four panels on the left-hand-side are at $t=2$ Gyr, while the four panels on the right-hand-side are at $t=3$ Gyr.\label{fig:morphologyy_seeding}
        }
\end{figure*} 
We first focus on the most important morphological structures in the simulations, i.e. face-on and edge-on slices of the density and the magnetic field strength for the Milky Way-like models MW-snB and MW-primB. The slices of the face-on view are centred in the mid-plane, the ones for the edge-on view are perpendicular to it. Figure \ref{fig:morphologyy_seeding} shows our results for the model MW-snB after $t=2$ Gyr (first and second columns) and $t=3$ Gyr (third and fourth columns). The upper left panel shows that the gas in the disc builds up a well defined spiral structure. The lower left panel indicates that the system evolved to a thin disc that is surrounded by the lower density gas of the CGM. However, because of the density gradient of the underlying $\beta$-profile one can see a brighter blue halo closer to the disc. This density gradient is a quantity which has been introduced by our galactic model itself and is only weakly influenced by the boundary layer between the gas disc and the CGM.
In the second panel on the top, we show the face-on view of the absolute magnetic field strength. While the gas density on average decreases monotonically towards the outer parts of the disc, the situation is different for the galactic magnetic field which is of the order of a few $\mu$G in two different regions. This is in good agreement with observations of several spiral galaxies \citep{rbeck13, Beck2015}.
The first region with strong magnetic fields is the very centre of the galactic disc (i.e. the innermost kpc). The second region is a ring-shaped structure further outside in the galactic disc and approximately located at $10$ kpc distance from the galactic centre. The former originates from the large SFR in the very centre of the galaxy. A consequence of strong star formation is a large rate of supernovae in the same region. In the model MW-snB this is directly correlated with the amount of magnetic dipoles that are seeded into the surrounding ISM, leading to a strong magnetic field. Furthermore, in this regime the magnetic field is exponentially amplified by the small scale turbulent motion of the ISM. This small scale turbulence is mainly driven by the higher supernova rate in the centre of the galaxy. We will discuss this in more detail in section \ref{sec:amplification}.
In contrast, the amplification process in the outer parts of the disc is most likely not driven by small scale turbulence. There, we observe very high rotation velocities of the gas due to the differential rotation of the disc. This leads to an exponential amplification  of the magnetic field  through the kinematic motion known as the $\alpha-\omega$-dynamo. This explains the strong magnetic field at the edge of the disc, although the amount of supernovae in the ambient medium in the outer parts of the disc is significantly lower compared to the galactic centre. Nevertheless, turbulent motion could also play a crucial role in this regime.

\begin{figure*}
        \includegraphics[scale=0.9]{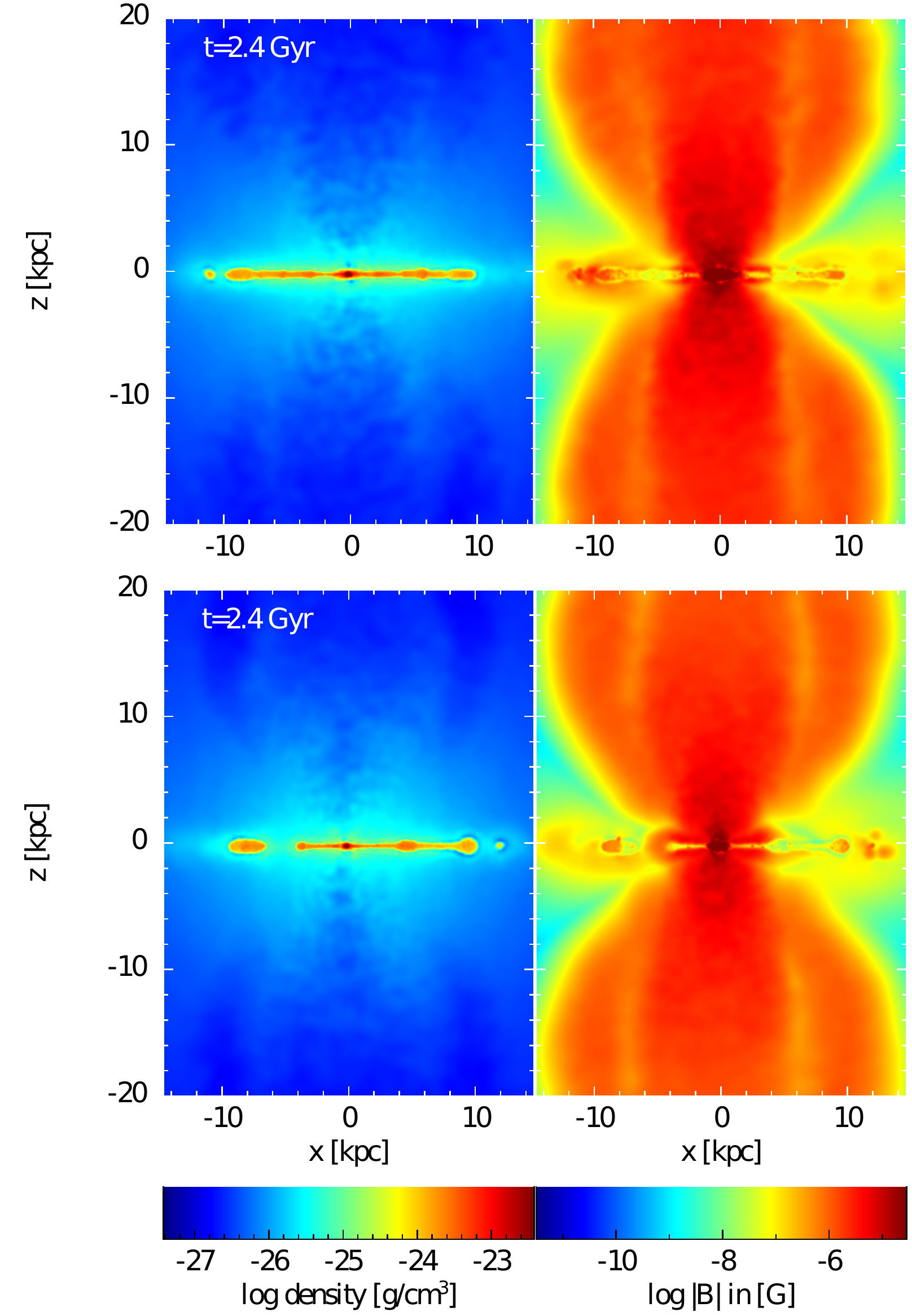}
        \caption{Cross section slices for the \textit{MW-snB} run (top panels) and the \textit{MW-primB} run (bottom panels) at $t=2.4$ Gyr. On the left we show the edge-on view of the gas density and on the right we show the edge-on view of the magnetic field strength. For both models we observe a strong biconical outflow in the magnetic field strength with an x-shaped structure.
\label{fig:xshape}}
\end{figure*}

\begin{figure*}
        \centering
        \includegraphics[scale=0.55]{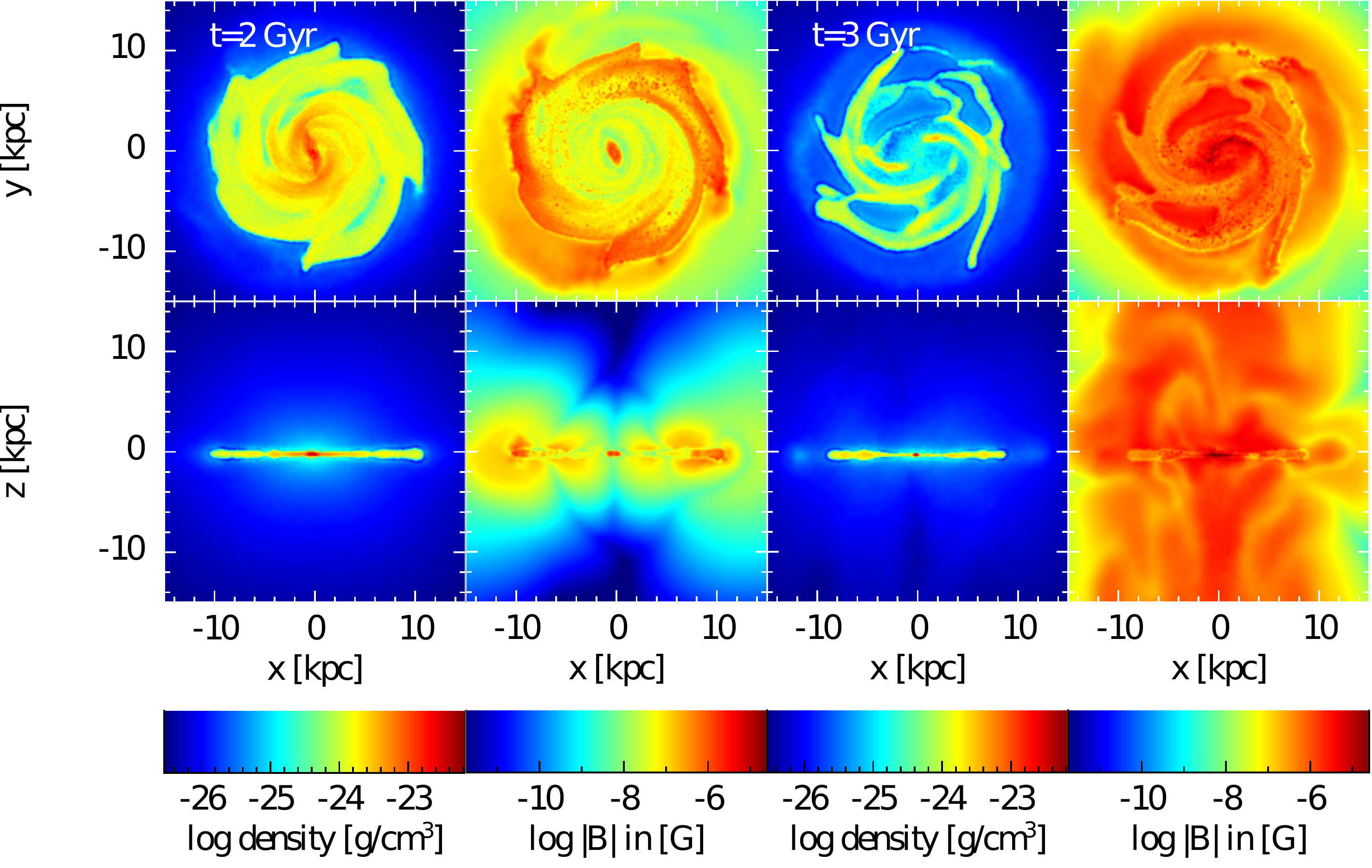}
        \caption{Same as figure \ref{fig:morphologyy_seeding} but for the \textit{MW-primB} simulations. A similar behaviour as in figure \ref{fig:morphologyy_seeding} is observed.
        \label{fig:morphologyy_Bx}
        }
\end{figure*}

\begin{figure*}
        \centering
        \includegraphics[scale=0.45]{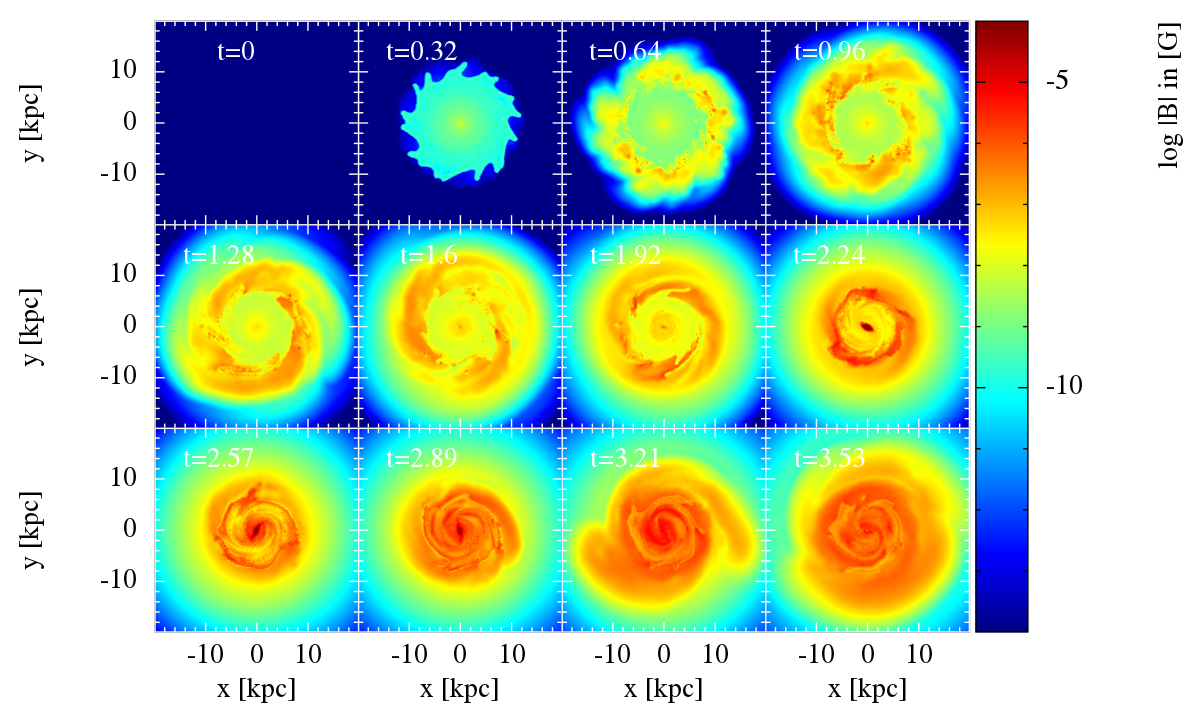}
        \includegraphics[scale=0.45]{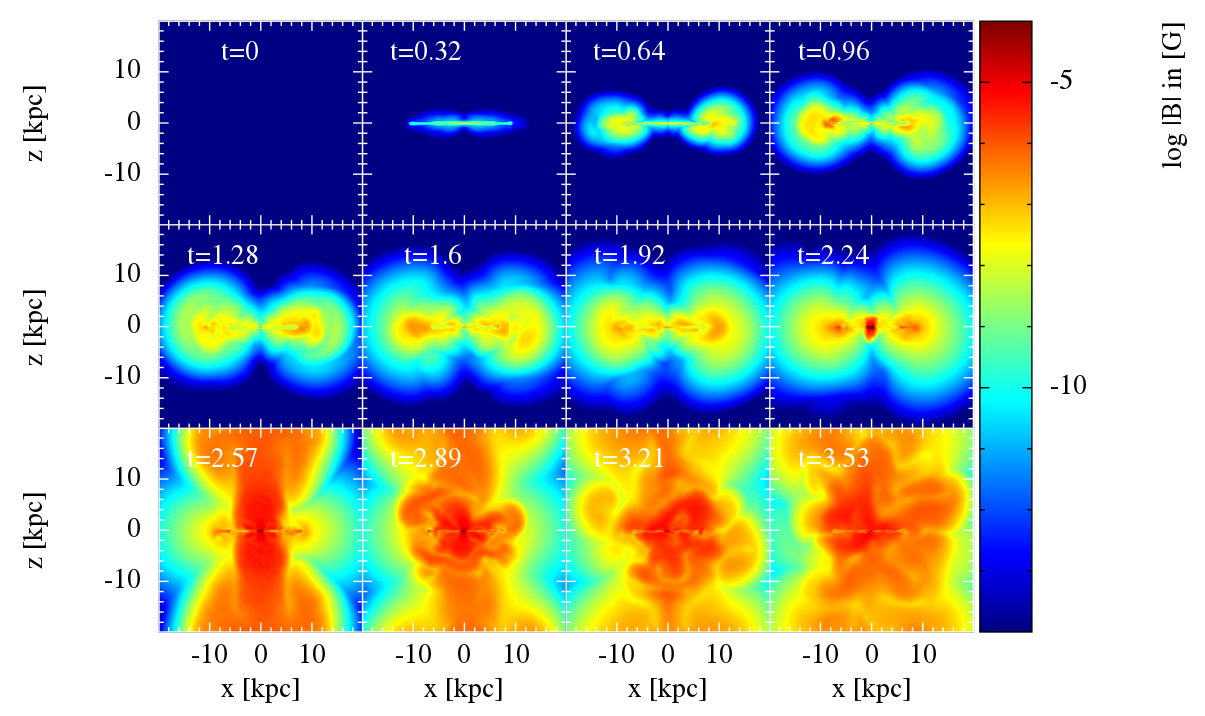}
        \caption{Face-on (top panels) and edge-on (bottom panels) time sequences for the simulation \textit{MW-snB}. The sequence shows the amplification of the magnetic field strength in the galactic disc, as well as the development of the outflow of highly magnetized material perpendicular to the disc.}
        \label{fig:seq}
\end{figure*}

We note that the large magnetic field strengths are correlated with the location of the spiral arms in the gas density, but they do not explicitly follow each other. It seems that the magnetic field strength between or at the edges of the spiral arms in the gas density is higher than in the gaseous spiral arms itself. This behaviour is indicated by observations (see \citet{Beck2015} for a review) and can be explained in galaxies with high density. Strong density waves lead to a compression of the magnetic field at the inner edges of the spiral arms. This can lead to turbulent motion and further amplification of the magnetic field. However, on
e can hardly see whether the magnetic field follows the spiral structure of the gas density by only comparing the two top left panels of figure \ref{fig:morphologyy_seeding} at this point in time. The anti-corellation becomes clearer later at $t=3$.
We will evaluate this behaviour in closer detail in section \ref{sec:magstruct_sec}.
In the bottom row of figure \ref{fig:morphologyy_seeding} we plot the edge-on view of the gas density (first panel) and the magnetic field strength (second panel).
Perpendicular to the galactic disc, the magnetic field strength declines much more weakly than the gas density.
This is especially true at the edge and is a result of the interaction between the outer parts of the disc with the surrounding CGM. The large magnetic field at the edge leads to a weak magnetization of the nearby
CGM. Due to the magnetic pressure the gas particles near the edge of the disc gain momentum in z-direction, move into the CGM, slow down, magnetize the nearby particles in the CGM, stop and fall back towards the disc. 
After $t=3$ Gyr (third and fourth columns), the face-on view of the gas density shows that the spiral structure is more depleted than at $t=2$ Gyr. We note that this effect is the strongest in the galactic centre. This can be explained by the higher SFR in the centre of the galaxy. A large fraction of the original gas mass in the centre has been converted into stars which leads to a significant drop in the gas density.
In the third panel on the top, we present the face-on view of the magnetic field for $t=3$ Gyr. To this point in time the magnetic field strength in the galactic disc has significantly increased and evolved towards a prominent spiral structure. It reaches values of a few $10 \mu$G in the magnetic spiral arms to a few $100 \mu$G in the galactic centre. The magnetic field in the spiral arms is still in agreement with the observations of magnetic fields in spiral galaxies \citep[e.g.][]{rbeck13}. The magnetic field in the central region has a maximum value that is around $350 \mu$G. Although, this is a high magnetic field strength there is observational evidence that the magnetic field in the centre reaches values between $50 \mu$G \citep{Crocker2010} and $4$ mG \citep{Yusef1996}.

We further note that at $t=3$ Gyr, we can clearly observe the magnetic field in the inter-arm regions to be stronger than in the gaseous spiral arms.
We are also able to reproduce the correct field strengths in the inter arm regions of a few $10 \mu$G that are known from observations \citep{Beck2007, Beck2009}. Our simulations reach larger field strengths in the inter-arm regions than in the gaseous spiral arms with the correct field strengths compared to observations. However, the physical reason for this behaviour cannot be clearly determined.
There are several possibilities that can lead to stronger magnetic fields in the inter-arm regions, like the buoyancy, Parker, or magnetorotational instabilities \citep{Beck2015}. 

In the centre of the galactic disc the magnetic field is further amplified by the small scale turbulent motion of the ISM. This leads to a very huge magnetic pressure in the very innermost kpc of the galactic disc. When the magnetic pressure becomes large enough it can accelerate particles alongside the z-direction. The strong magnetic field drives the system out of equilibrium, leading to a sharp pressure gradient between the magnetic dominated disc and the thermal dominated, hot gaseous halo. As a result the magnetic field lines break up in z-direction to reduce the sharp pressure gradient at the edge of the disc and we can observe a biconical outflow of magnetic energy perpendicular to the disc. We note that it has its origin in the very centre of the disc.

Because this biconical magnetic tube is one of the central morphological features we show it for both magnetic field models (\textit{MW-snB} and \textit{MW-primB}) in figure \ref{fig:xshape} shortly after it sets in at $t=2.3$ Gyr. The top row of figure \ref{fig:xshape} shows the edge-on
gas density (left) and magnetic field (right) for the \textit{MW-snB} run. The bottom row shows the same quantities for the \textit{MW-primB} run. 
Although magnetic fields are seeded differently in both models,
we observe this biconical magnetic tube in both simulations at the exact same time. In the beginning the magnetic tube rises symmetrically above the disc. It moves forward into the outer parts of the CGM, with a mean velocity of around $400-500$ km/s. The biconical tube transports a significant amount of magnetic energy into the CGM, leading to its magnetization.

\begin{figure}
	\includegraphics[scale=0.37]{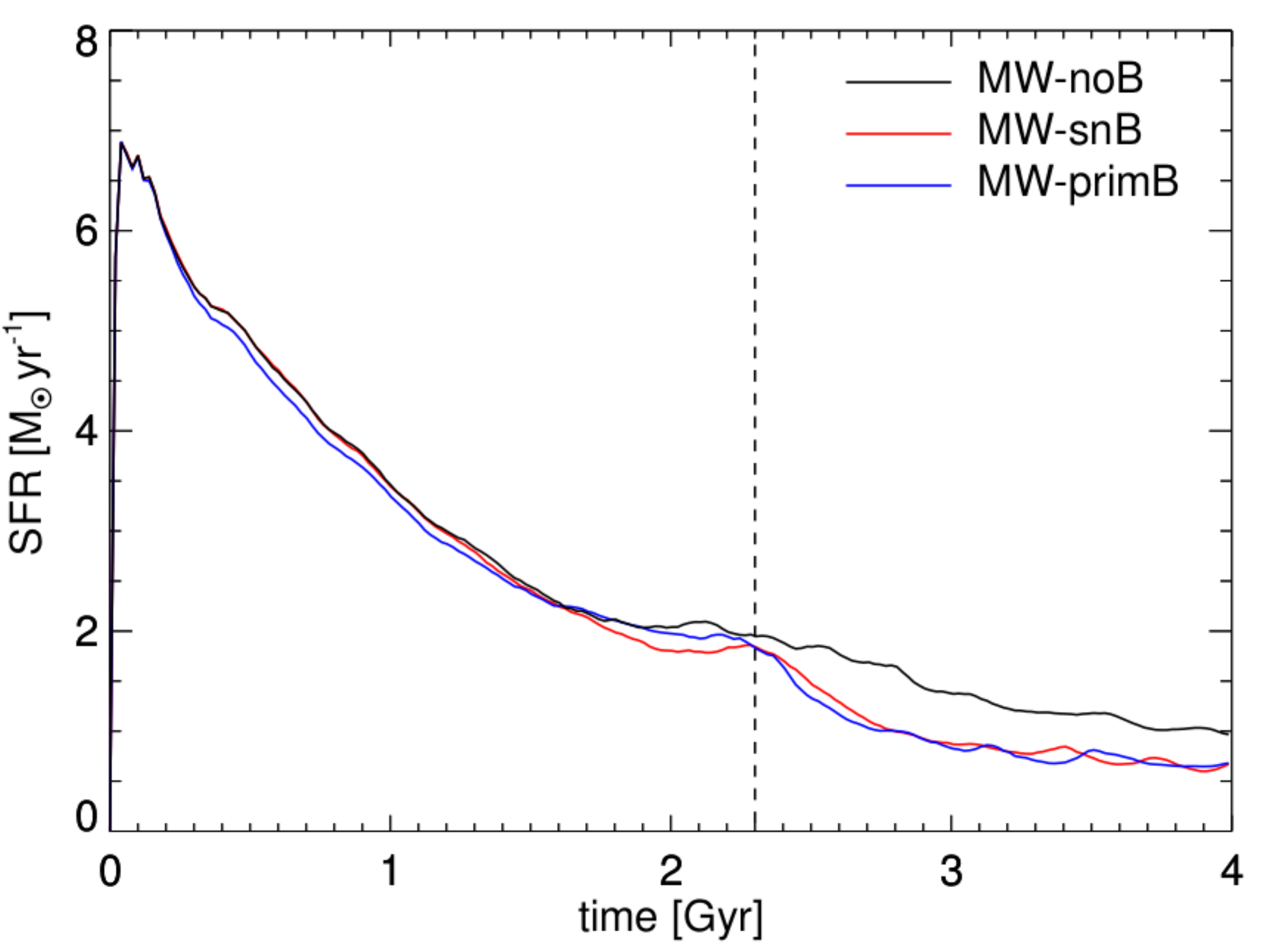}
        \caption{SFRs for the simulations \textit{MW-noB} (black line), \textit{MW-primB} (red line), and \textit{MW-primB} (blue line). For all three runs the SFR peaks shortly after the simulation starts due to the non-equilibrium initial condition. At $t=2.3$ Gyr the SFR drops for both models that include magnetic fields. \label{fig:sfr}}
\end{figure}

Moreover, this effect leads to morphological features in the gas density of the CGM close to the disc at the onset of the biconical magnetic tube at around $t=2.3$ Gyr.
The nearby CGM shows a X-shaped or H-shaped structure. These structures of the CGM around galaxies are a well known morphological feature in observations of galaxies with an active galactic nucleus (AGN) in the centre \citep[i.e.][]{Veilleux2005}. In these observations it is known as an indicator for a biconical galactic outflow. We note that we do not include such an AGN, but we observe the CGM structures that indicate such an outflow. In our simulations this outflow would be driven by the magnetic field. Therefore, we discuss the possibility of magnetic driven biconical outflows in more detail in section \ref{sec:CGMandDisk}. We note that we observe these X-shaped structures of the CGM around the galaxy only in the first few $100$ Myr after the onset of the biconical magnetic tube and thus we conclude that the magnetic wind should be the strongest shortly after it sets in.
We do not find a huge difference in the X-shaped structures around the galaxy between the models \textit{MW-snB} and \textit{MW-primB}. Nevertheless, we observe slightly higher magnetic field strengths in the biconical tube for the model \textit{MW-snB}. 
In figure \ref{fig:morphologyy_Bx} we present the exact same properties as in the figure \ref{fig:morphologyy_seeding} for the model \textit{MW-primB}. 
The overall morphological structure is comparable to the model \textit{MW-snB}. We observe a prominent gas disc in the face-on view for $t=2$ Gyrs as well as a thin gas disc in the edge-on view. Further, we see that the magnetic field is mainly amplified in the galactic centre and in a ring around the centre with a radius of around $10$ kpc. 
As we do not seed magnetic field with supernovae, we emphasize 
that the amplification of the magnetic field in
the model \textit{MW-snB} is not caused by a higher magnetic field seeding in the central region due to the higher amount of supernovae.
In both cases the supernovae are a crucial component for the amplification of the magnetic field in the galactic centre. But the cause for the amplification is the turbulent motion that is driven by supernova-feedback in the galactic centre leading to an exponential amplification of the magnetic field in the centre. By comparing models \textit{MW-snB} and \textit{MW-primB} it becomes clear that turbulence induced by supernova-feedback is the crucial component for the amplification of the magnetic field in the centre.
We present more evidence of the small-scale turbulent dynamo in section \ref{sec:amplification}. Further, our results justify the common choice of a primordial seed field which has often been used by other groups \citep[e.g.][]{Pakmor2013, Butsky2017}.

\begin{figure}
        \includegraphics[scale=0.37]{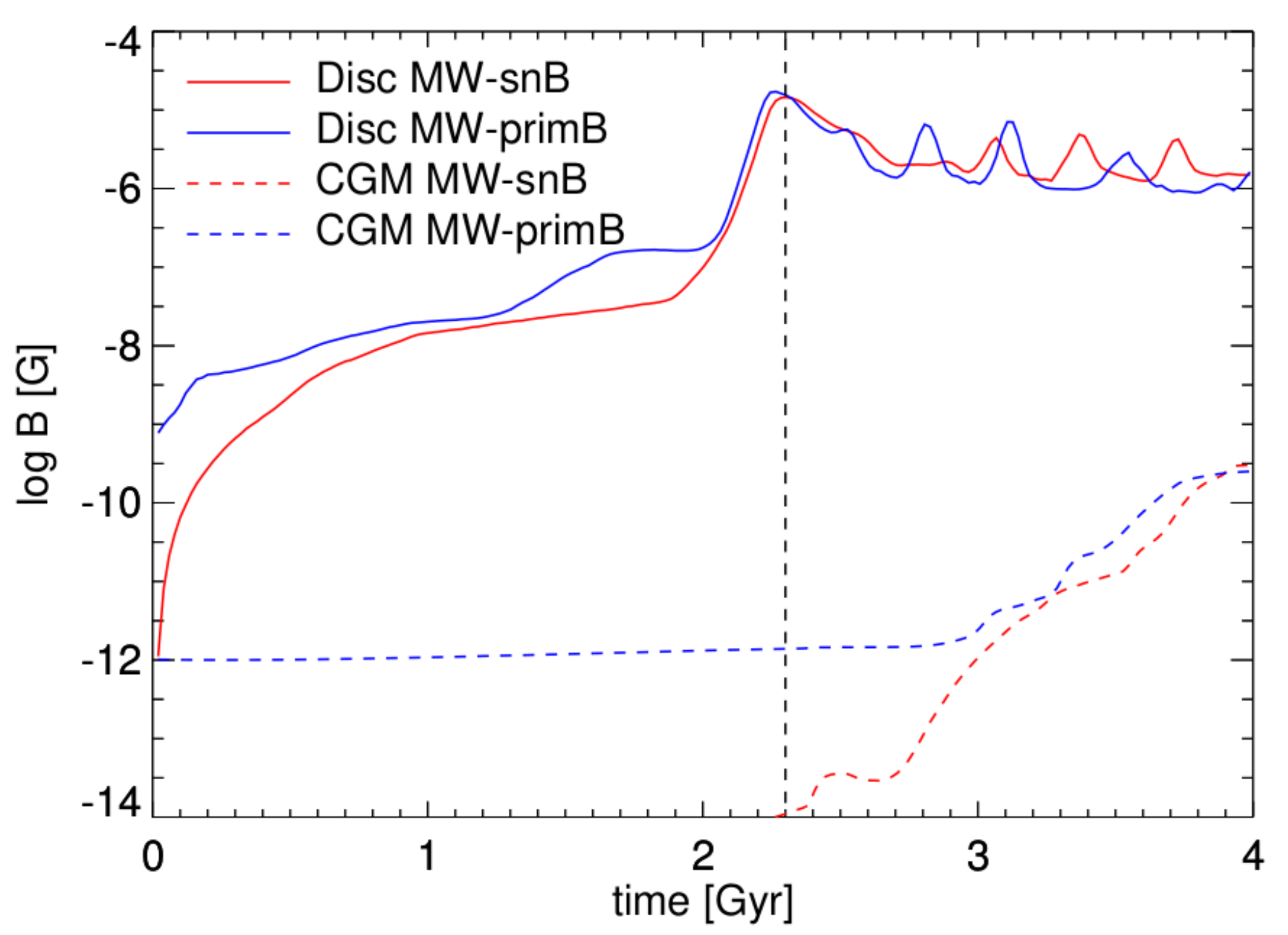}
        \caption{Magnetic field strength for the simulations
        \textit{MW-snB} (red line) and \textit{MW-primB} (blue line).
        In the disc, it first growth exponentially and then
        saturates at $\sim10^{-6}$ G with oscillations that reach up to a few $10^{-5}$ G. The dashed lines show the evolution of magnetic field strength in the CGM. The mean magnetic field strength in the CGM is around $10^{-10}$ G and is still rising when the simulation stops. \label{fig:magneticgrowthrate}}
\end{figure} 

The edge-on view (bottom panels of figure \ref{fig:morphologyy_Bx}) shows the same biconical magnetic outflow. Its appearance in the model \textit{MW-primB} leads to the conclusion that this structure in the magnetic field is not driven by the supernova-seeding. Furthermore, we observe a similar behaviour of the magnetic field in the inter-arm regions which becomes even clearer for the model \textit{MW-primB} at $t=3$ Gyr.
The model \textit{MW-primB} is also capable of reproducing the observed field strengths in spiral galaxies. To give a more detailed overview of the evolution of the magnetic field, we show a time sequence of the evolution of the magnetic field strength for the model MW-snB in figure \ref{fig:seq}.
The top panels show the face-on view, and the bottom panels show the edge-on view.
The evolution of the magnetic field in the \textit{MW-primB} model is very similar, with only minor differences.

\subsection{Star formation rate within the different models \label{sec:star formation}}

In figure \ref{fig:sfr} we show the SFR for the simulations \textit{MW-snB}, \textit{MW-primB} and \textit{MW-noB} as a function of time.
All simulations were evolved for $t=4$ Gyr. The SFR at the beginning of the simulation is very similar in all three runs, and peaks shortly after the start of the simulation to a value of around $7 M_{\odot}/yr$.
The initial peak can be explained by the non-equilibrium configuration and the large amount of cold gas at the start of the simulation. 
It is entirely set by the initial conditions, and does not depend on the magnetic field model. After the initial peak, the SFR declines rapidly. In the first $1.5$ Gyr, there is no significant difference between the three models, except for
a slightly lower SFR for the \textit{MW-primB} run. This small deviation can simply be explained by the difference in the initial magnetic field strength.
While the simulations \textit{MW-snB} and \textit{MW-noB} start without any magnetic field the simulation \textit{MW-primB} starts with a field strength of $10^{-9}$ G in the disc. This is demonstrated in figure \ref{fig:magneticgrowthrate}, which shows the magnetic field strength for the simulations \textit{MW-snB} (red solid line) and \textit{MW-primB} (blue solid line). 
The mean magnetic field strength of the \textit{MW-primB} run is three orders of magnitude higher in the beginning than in the \textit{MW-snB} run, leading to a higher magnetic pressure, which can results in a lower SFR if gas is pushed out of the star-forming regions of the disc.
The SFRs in the simulations \textit{MW-snB} and \textit{MW-noB} are identical in the beginning, since the small initial
magnetic field corresponds to the run without a magnetic field
and therefore does not change the SFR. After the SFR drops, it remains nearly constant between $1.7$ Gyr and $2.3$ Gyr for all three simulations.
Then the SFR in the simulations with magnetic fields drops by about $50$ per cent for the rest of the simulation compared to the simulation without magnetic fields.
This indicates that in the \textit{MW-snB} and \textit{MW-primB} runs, a significant amount of the star-forming gas is removed from the disc. This can be explained by the results of section \ref{sec:morphology}, which showed
strong magnetized biconical outflows that set in at around $t=2.3$ Gyr
in both runs. This indicates that the SFR is reduced due to a magnetized outflow of gas from the disc to the CGM. 
This outflow is driven by the magnetic pressure, i.e. the magnetic field strength within the disc, that rises mainly in the centre due to amplification via small scale turbulence. This results in a smaller gas reservoir in the disc leading to a lowered SFR. In the star-formation model used in our simulations, all gas above the threshold density forms stars, independent of its temperature and magnetic field strength. 
Thus, we cannot follow the impact of the magnetic field on the SFR directly, but only its indirect influence, such as outflows that are driven by the contribution of the magnetic pressure in the ISM.
For a more detailed analysis on how the magnetic field influences the galactic SFR, the density threshold in the star formation recipe would have to be changed to a pressure threshold, such that the magnetic pressure $\mathbf{B}^2/8\pi$ can be taken into account directly.

\subsection{Amplification of the magnetic field \label{sec:amplification}}

\begin{figure*}
\vspace{-6.5cm}
\makebox[\textwidth]{
\hspace{0.5cm}
\includegraphics[width=1.3\textwidth]{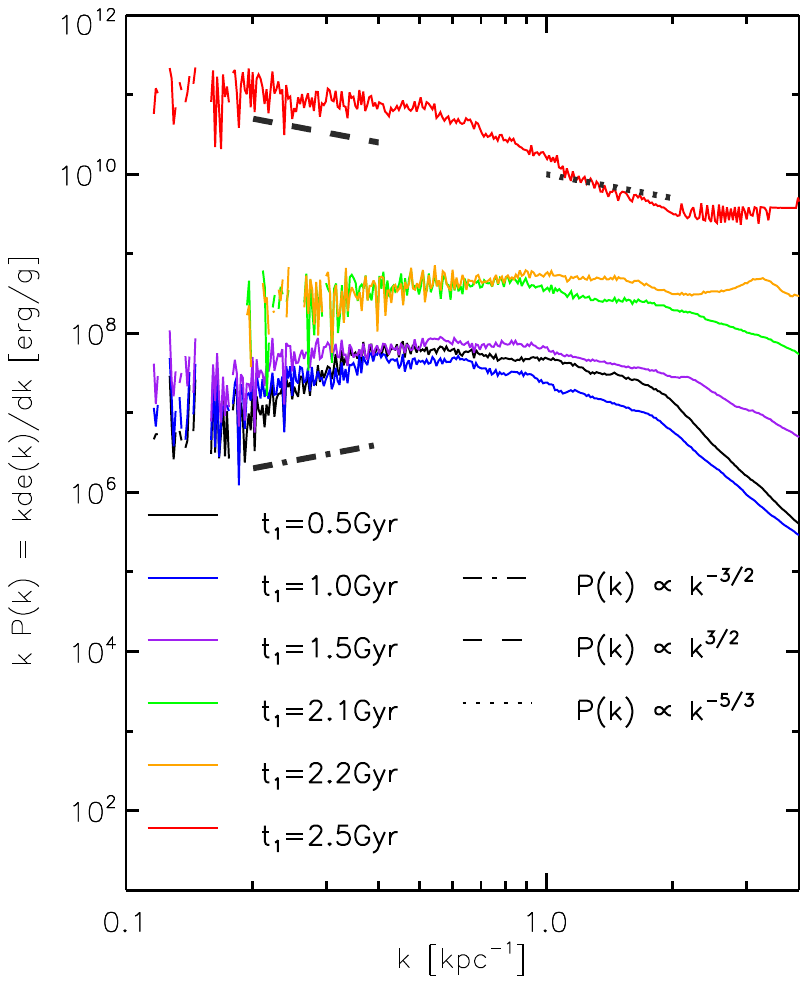}
\hspace{-14.5cm}
\includegraphics[width=1.3\textwidth]{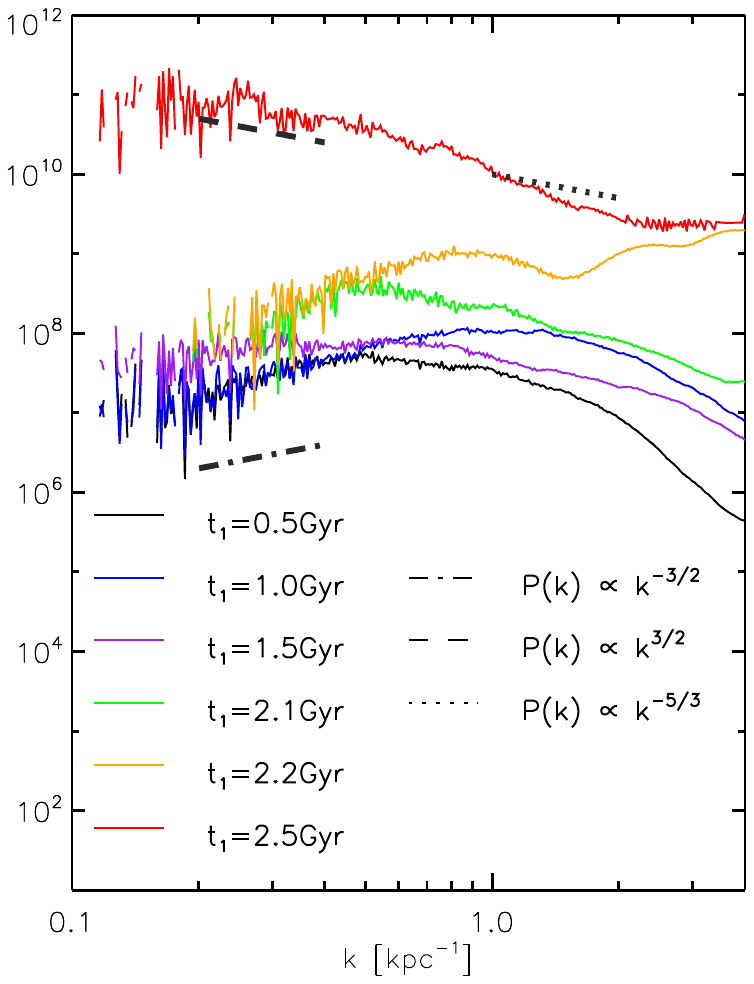}
}        
\vspace{-15.5cm}
\caption{Magnetic power spectra for the simulations \textit{MW-snB} (left-hand-side panel) and MW-primB (right-hand-side panel). 
Both models result in very similar power spectra, independent of the seeding model
The magnetic field is amplified by turbulent motion on small scales, which
is transported to large scales through an inverse energy cascade. This \citet{Kazantsev1968} spectrum is an indicator for a small scale turbulent dynamo, resulting in an increase of the power $P(k) \propto k^{3/2}$ on large scales. The small scale dynamo stops at later times due to the large magnetic field resulting in
an \citet{Iroshnikov1963} spectrum with $P(k) \propto k^{-3/2}$.}
\label{fig:power}
\end{figure*}

\begin{figure}
\vspace{-2.5cm}
\hspace{-4cm}
\includegraphics[scale=0.6]{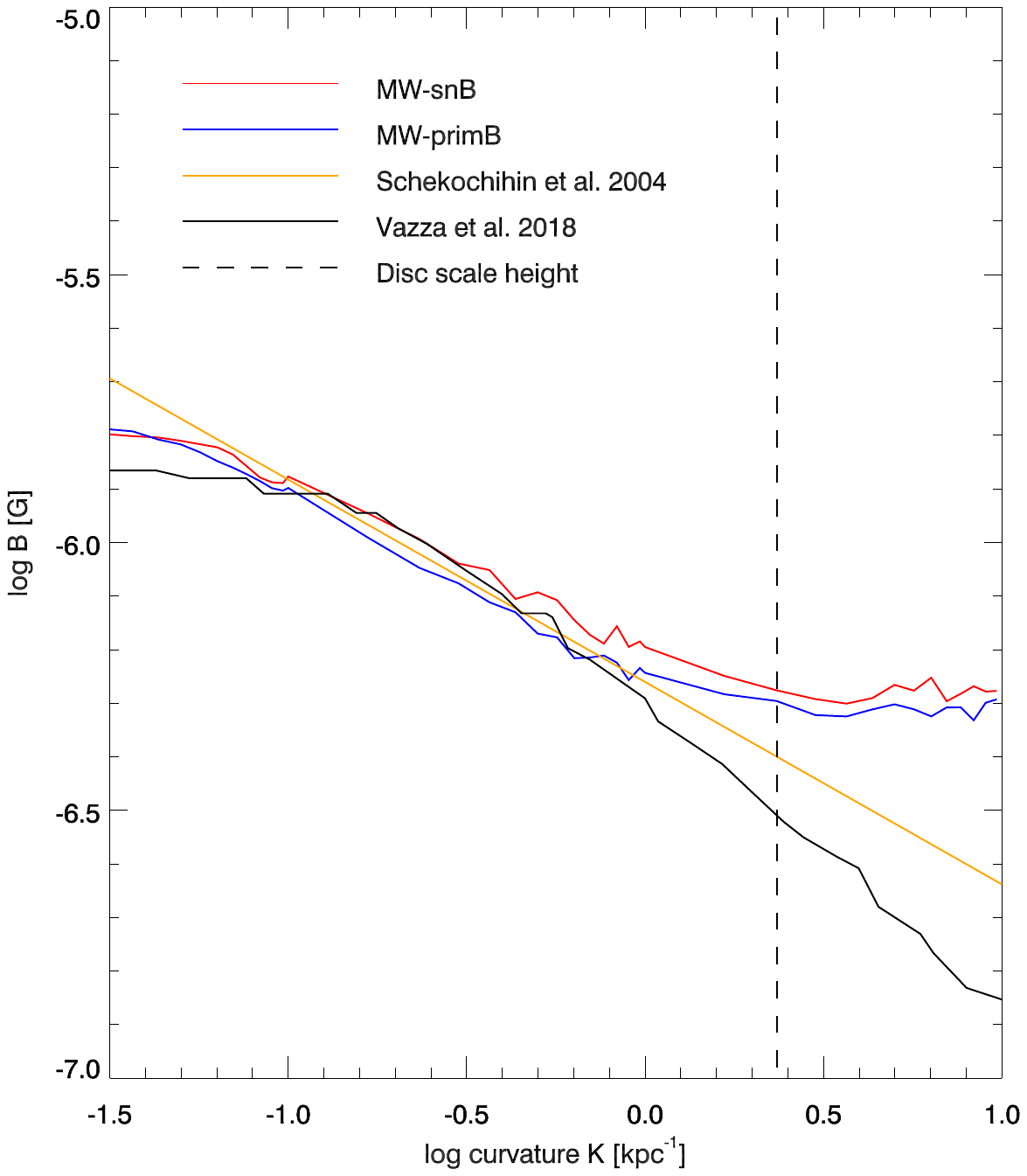}
\vspace{-2cm}
\caption{Median of the magnetic field strength as a function of the curvature of the magnetic field lines for $t=2$ Gyr. The red line
represents the \textit{MW-snB} run and the blue line the \textit{MW-primB} run.
The field strength and curvature are anti-correlated with a power law slope of $0.5$, as the field lines are harder to bend in the presence of stronger magnetic fields.
The slope indicates that the amplification is driven by small scale turbulence in agreement with the results by \citet{Schekochihin2004} (orange line). The black line shows the recent result of MHD-simulations of galaxy-clusters by \citet{Vazza2018}. \label{fig:curvature}}
\end{figure}

A very important aspect of disc galaxy simulations with SPMHD is to reproduce the observed magnetic field strengths in the disc. 
There are many observations of magnetic fields in galactic discs \citep[e.g.][]{Hummel1986, Chyzy2003, Chyzy2007, Beck2007},
which show that the field strength in discs ranges from $10 \mu G$ between the spiral arms up to $50 \mu G$ within the spiral arms. We present the growth rate of the magnetic field in our three models in figure \ref{fig:magneticgrowthrate}. The solid lines represent the magnetic field strength in the galactic disc, for the models with supernova-seeding (red) and a constant magnetic seed-field (blue). The dashed lines represent the magnetic field strength in the CGM. The vertical dashed line marks the point in time when the magnetic pressure becomes dynamically important at around $t=2.3$ Gyr.
At the beginning of the simulation the magnetic field in the disc roughly grows exponentially in both models, in good agreement with the findings of dynamo theory. 
In galaxies there are in general two amplification processes for the magnetic field. 
The first is the small scale turbulent dynamo and the second is the mean-field $\alpha$-$\omega$-dynamo. Both dynamos can lead to either exponential or linear growth of the magnetic field. In the case of the small scale turbulent dynamo, the amplification of the magnetic field happens due to the turbulent motion in the ISM as long as we can neglect the pressure caused by the magnetic field itself, so that the dynamo operates in the kinetic regime \citep{Pakmor2017}. The magnetic energy rises exponentially until an equilibrium with the kinetic energy is reached. At this point the magnetic energy can be transported to large scales due to an inverse energy cascade. In this regime the small scale dynamo is only able to follow linear amplification \citep{Federrath2016}. In the case of the $\alpha$-$\omega$-dynamo, the differential rotation and the $\alpha$-effect (small scale vertical motion of the gas particles) in the galactic disc itself
can lead to both, exponential or linear growth of the magnetic field. 
It is not clear which of those two amplification process is favoured in our simulations. However, we can deduce the dominant amplification process from the power spectrum of the magnetic field, which is shown for both models in figure \ref{fig:power}.
These power spectra have been obtained with the tool \textsc{sphmapper} \citep{Roettgers2018}, which carries out an appropriate binning for SPH data on a regular grid using the same kernel that is used in the simulations (Wendland C4).
The left panel of figure \ref{fig:power} shows the power spectra at five different points in time for the model \textit{MW-snB}. The right panel of figure \ref{fig:power} shows the power spectra at the same points in time for the models \textit{MW-primB} field. Both magnetic field models reproduce a relatively smooth distribution of the magnetic power from small scales to large scales.
We find strong evidence for a small scale turbulent dynamo for both magnetic field models. From dynamo theory we expect a power spectrum $P(k) \propto k^{3/2}$ in the case of a small scale turbulent dynamo. We can see this behaviour in both power spectra very clearly, especially at the beginning of the simulation, when the equilibrium state between the magnetic and the kinetic energy has not been reached yet. The magnetic field is amplified by turbulent motion on small scales and transported to large scales by an inverse energy cascade, as predicted by dynamo theory. We note, that the power spectrum on the large scales is fully consistent with a Kazantsev-spectrum \citep{Kazantsev1968, Kraichnan1968}, known from a small scale turbulent dynamo \citep{Brandenburg2005, Tobias2011}. This is also consistent with the findings of other simulations of isolated disc galaxies \citep{Butsky2017, Rieder2016, Rieder2017a}, as well as those of cosmological zoom-in simulations \citep{Pakmor2017, Rieder2017b}.
As this small scale dynamo is one of the central findings of our study, we 
provide more evidence based on the \citet{Kazantsev1968} theory. For this,
we calculate the magnetic curvature \textbf{K} given by \citet{Schekochihin2004}
\begin{align}
  \textbf{K} = \frac{\left( \textbf{B} \cdot \nabla\right) \textbf{B}}{|\textbf{B}^{2}|}.
  \label{eq:curvature1}
\end{align}
By using vector identities we can reformulate equation \ref{eq:curvature1}. Thus we obtain
\begin{align}
  \textbf{K} = \frac{1}{|\textbf{B}^2|}\left[\frac{1}{2} \nabla \left( \textbf{B} \cdot \textbf{B} \right) - \textbf{B}\times \left(\nabla \times \textbf{B} \right) \right].
\end{align}
The magnetic curvature can be used to distinguish the regime where the magnetic field is amplified by adiabatic compression (i.e. the magnetic field strength stays constant with increasing curvature) and the regime where a dynamo is acting \citep{Schekochihin2004, Schober2015}. In this case, an anti-correlation can be found
between the magnetic field strength and the curvature with the relation $KB^{0.5} = const.$ We show the median magnetic field strength as a function of curvature in figure \ref{fig:curvature}. The red and blue lines corresponds to the \textit{MW-snB} and \textit{MW-primB} runs, respectively. The orange line represents the power law slope indicated by \citet{Schekochihin2004}. Further, the black line shows the recent results by \citet{Vazza2018}, where the same power law behaviour is found for a galaxy cluster simulated with the grid code \textsc{enzo}.
In our simulations, we find the same power law behaviour as \citet{Schekochihin2004}
for the intermediate curvatures between $0.1$ and $1$, which provides further evidence for a small scale dynamo in both our magnetic field models. Moreover, in figure \ref{fig:curvature} we indicate the curvature corresponding to the
disc scale height with a vertical dashed line. While in a cluster environment
the power law slope of \citet{Schekochihin2004} can be recovered also in the high curvature regime, this is not the case for our isolated galactic systems. The reason for this may be that the interface between the rotating disc and the halo gives a natural scale on which the field has to be bend. We note that the curvature is a very noisy quantity, which has two origins.
The first one is due to  the rapidly changing distributions of both the magnetic field strength, and the curvature in agreement with \citet{Schekochihin2004}.
The second one can be explained by the low order gradient estimates that we used for calculating the curvature as a real SPH-quantity \citep[e.g.][]{price12}.

Altough, we can present strong evidence for a small scale turbulent dynamo in the power spectra (figure \ref{fig:power}) 
and the magnetic curvaturea (figure \ref{fig:curvature}),
at later times, the slope of the power spectrum is no longer in agreement
with the \citet{Kazantsev1968} power spectrum anymore. This indicates
that the amplification process is not dominated by the small scale dynamo at later times. In this case we find that our power spectra  are similar to an \citet{Iroshnikov1963, Iroshnikov1964} spectrum that is acting in the regime of strong magnetic fields. Examining this behaviour we believe that the small scale dynamo is turned off at later times due to the strong dominating magnetic field in the galaxy. This is in agreement with the behaviour we observed in figure \ref{fig:magneticgrowthrate}, where we see an exponential growth in the beginning, which becomes linear at later times. The interplay between the Iroshnikov-spectrum and the linear growth of the magnetic field at later times leads to the conclusion that the amplification process of the magnetic field in this regime is either driven by the $\alpha$-$\omega$-dynamo instead of the small scale turbulent dynamo, or switched off completely.

Finally, we note that in the CGM there is nearly no growth of the magnetic field visible. In the \textit{MW-primB} run, there is a small amplification of the magnetic field in the beginning, because of the none-zero magnetic field in the CGM in this model and the slight rotation of the CGM. However, the amplification in this case is minimal, and, as expected from dynamo theory, we can see a small exponential growth due to an $\alpha$-$\omega$-dynamo. Because the CGM is in hydrostatic equilibrium due to our initial conditions there is no amplification of the magnetic field strength over small scale turbulence.

Although there is no relevant magnetic field in the CGM in the first $2.3$ Gyr of the simulation, after around $2.3$ Gyr we see a jump in the magnetic field strength in the CGM of several orders of magnitude in both models.
There is no observable difference in the behaviour of the magnetic field strength in the CGM between both our models.
This underpins the fact that the observed magnetic field in our simulations depends mainly on the dynamical structure of the galaxy and is not dominated by the
seeding of the magnetic field.

\subsection{Halo accretion and magnetic driven ouflows \label{sec:CGMandDisk}}

\begin{figure}
\includegraphics[scale=0.37]{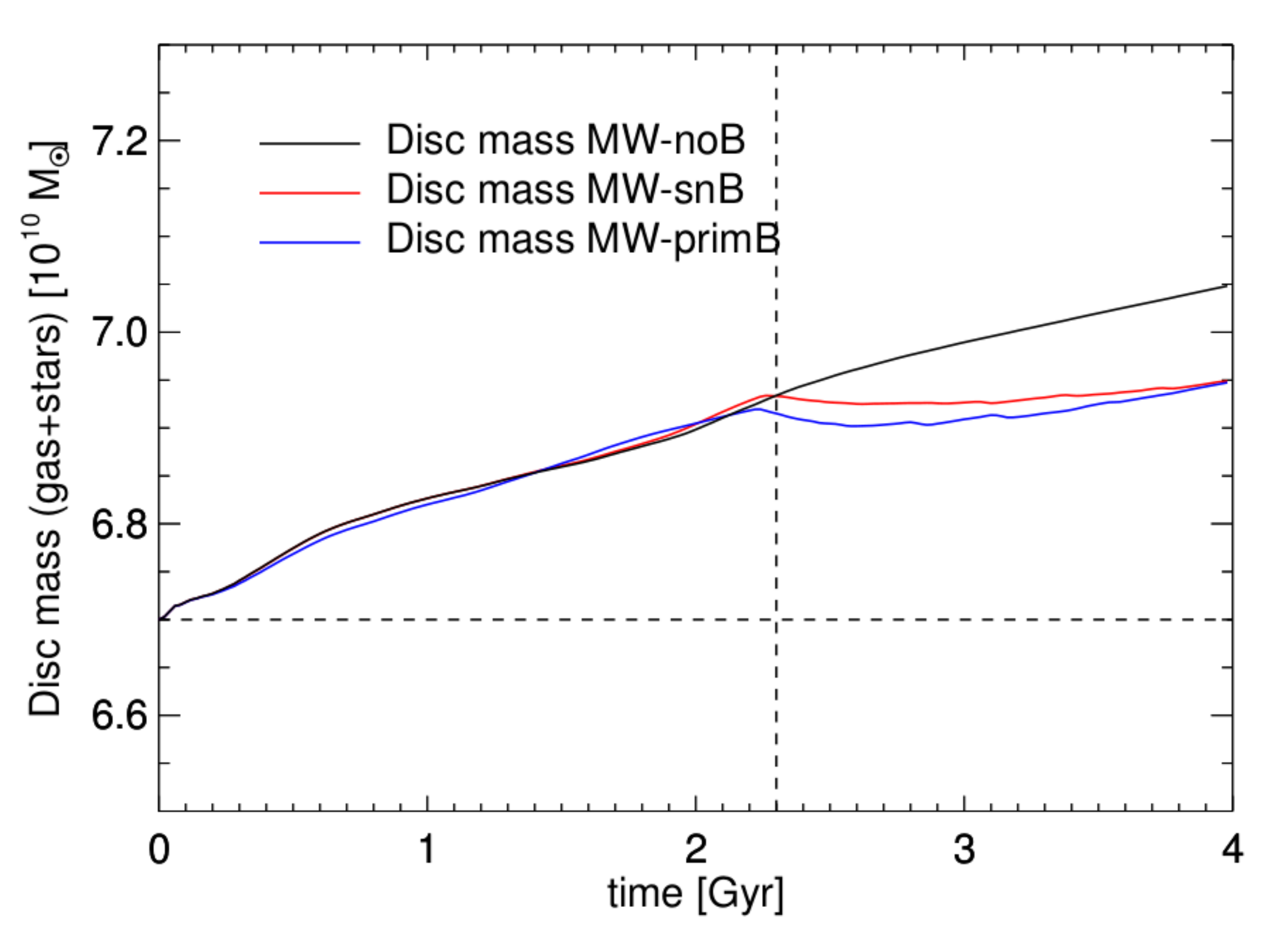}
\caption{Total mass of the disc for the simulations \textit{MW-noB} (black), \textit{MW-snB} (red), and \textit{MW-primB} (blue). While the disc is accreting a large amount of gas in \textit{MW-noB}, the net growth of the disc is suppressed in \textit{MW-snB} and \textit{MW-primB}, after the magnetic outflow sets in at $2.3$ Gyr.
\label{fig:coolingflow}}
\end{figure}

A new and very important aspect of our simulations is the inclusion of the CGM, unlike in previous simulations of isolated disc galaxies \citep[e.g.][]{Kotarba2011,Butsky2017}. This allows us to
observe the interaction of an isolated disc galaxy with its CGM in an idealised environment without any perturbations.   
In section \ref{sec:amplification}, we have shown that the CGM gets strongly
magnetized because of outflows 
that transport a lot of the magnetic power, which is amplified in the galactic disc via the small scale turbulent dynamo and the mean field $\alpha$-$\omega$ dynamo, to the outer parts of the CGM.
We observe a prominent tube with a radius of approximately $2$ kpc near the disc, which opens up to $5$ kpc in its outer parts with a magnetic field strength between $10^{-7}$-$10^{-6}$ G. It reaches the outer parts of the CGM with a total length of around $40$ kpc. This magnetic tube transports gas out of the disc in positive and negative z-direction at nearly the same rate, with a speed of a few $100$ km/s. Some of the gas particles are close to reaching the galactic escape velocity. The gas is moving outwards along the magnetized tube and falls back to the disc from outside of the tube, such that
there is an active exchange of the disc's gas with the hot gas of the CGM. Further, we note that after the onset of the magnetic tube we see heavily magnetized, low density bubbles rising from the galactic disc to the CGM reaching a hight of a few kpc above the disc, which is in agreement with the findings of \citet{Pakmor2013}.
In addition to the gas which is initially located in the disc, moves to the CGM, and falls back, there is also a cooling flow of hot gas from the CGM onto the disc.
To demonstrate this, we show the evolution of the total mass of the baryonic disc for our three models in figure \ref{fig:coolingflow}.
While cold gas is turned into stars, the total disc mass increases with time, as the disc is fuelled by cooling gas from the hot halo.
Interestingly, the cooling rate ($1{\rm~M}_{\odot}{\rm~yr}^{-1}$) is roughly of the same order as the SFR ($\sim3{\rm~M}_{\odot}{\rm~yr}^{-1}$), indicating, that the disc can compensate the loss of gas mass due to star formation by accreting hot gas from the CGM. Because of the immense gas reservoir, the SFR in the disc is thus stabilized, and eventually reaches an equilibrium.
After the magnetic outflow sets in, the total disc mass stays roughly constant, while in the simulation without magnetic fields the disc still grows by accreting gas from the CGM at a constant rate. Thus, for both magnetic field models the strong magnetic driven wind with about one solar mass per year, compensates the incoming cooling flow. This equilibrium between inflow and outflow results in a constant disc mass, although within the disc, cold gas is still converted into stars.
We note that the outflow velocties and mean magnetic field strengths in the biconal outflow we observe are failry similar to the properties observed in the 'Fermi-bubbles'.  In \citet{Carretti2013} a magnetic field strength between $6$ and $12 \mu$G is observed for the two lobes which is comparable to the field strenghts between $1$ and $30 \mu$G we find in the biconal outflow in our simulations. Furhter, in our simulations we find that these biconal structures propagate with a velocity of between $400$ and $500$ km/s into the CGM, while kinematic modelling of the Fermi bubbles leads to outflow velocities between $1000$ and $1300$ km/s as presented by \citet{Bordoloi2017}.   

\begin{figure}
\includegraphics[scale=0.37]{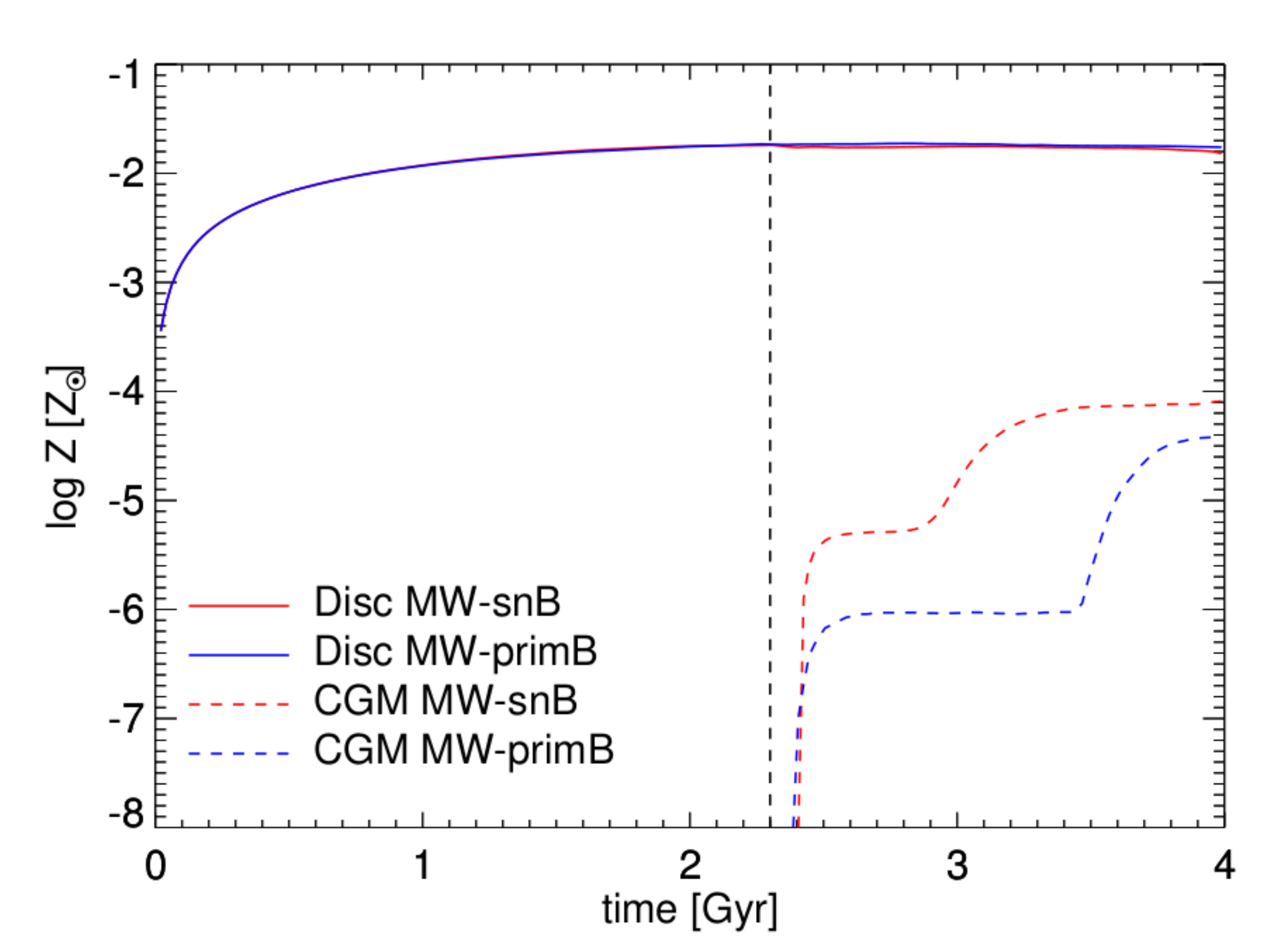}
\caption{Time evolution of metals in the disc (solid lines) and the CGM (dashed lines) for \textit{MW-snB} (red), and \textit{MW-primB} (blue). The black vertical dashed line indicates the onset of the outflow driven by magnetic fields, leading to a net magnetization of the CGM in both models that is normally believed to be obtained by the wind feedback induced by supernovae. \label{fig:metals}}
\end{figure}

Along with the outflowing gas, the metals which are returned to the disc by supernova feedback are transported to the CGM. We show this behavior for the models \textit{MW-snB} (red), and \textit{MW-primB} (blue) in figure \ref{fig:metals}. The solid lines represent the metals in the disc and the dashed lines the metals in the CGM. We observe metal enrichment in the CGM due to the magnetic outflow which is normally believed to be caused only by supernova driven winds from either late supernova driven winds in massive galaxies \citep[e.g.][]{Aguirre2001, Adelberger2003, Shen2012} or outflows from dwarf galaxies at higher redshift \citep[e.g.][]{Dekel1986, Maclow1999, Furlanetto2003}. Another mechansim for the metal enrichment of the CGM is proposed by \citet{Scannapieco2004} due to quasar driven winds. \citet{Gnedin1998} point out the importance of proto galaxy mergers at high redshift to enrich the surrounding medium with metals. Finally, we note that we could not observe any metal enrichment towards the CGM in the simulation \textit{MW-noB}.

\subsection{Magnetic field structure \label{sec:magstruct_sec}}

\begin{figure*}
\includegraphics[scale=0.4]{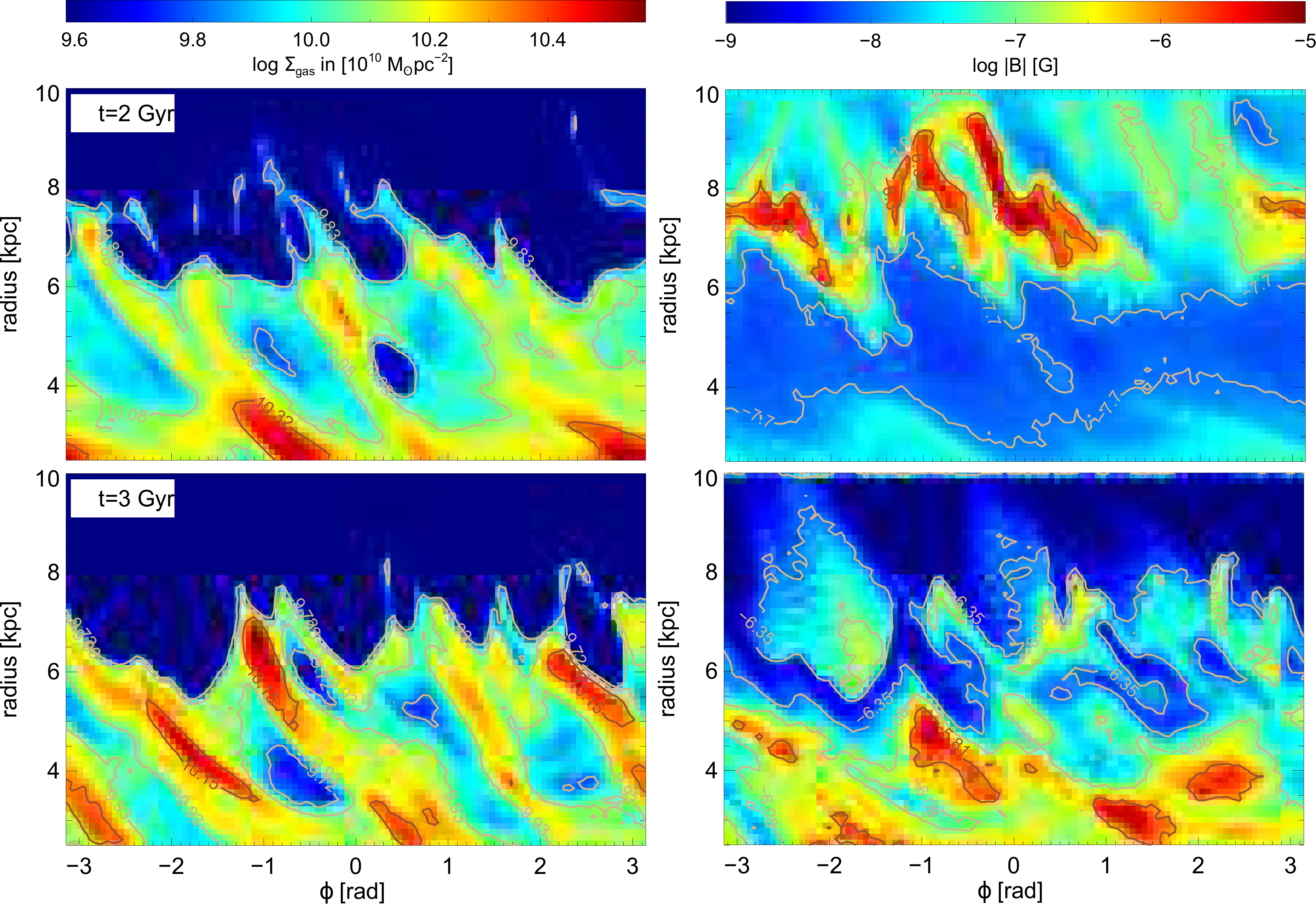}
\caption{Surface density of the gas (left-hand-side panels) and magnetic field strength (right-hand-side panels) in cylindrical coordinates for the \textit{MW-snB} run at 2Gyr (top panels) and 3 Gyr (bottom panels). The spiral structures in the magnetic field and the gas surface density are correlated in the centre at $t=3$ Gyr, while they are not
at $t=2$ Gyr. In the outer regions of the disc the magnetic field and the gas density are anti-correlated at $t=3$ Gyr.}
\label{fig:rphi1}
\end{figure*}

\begin{figure*}
\includegraphics[scale=0.4]{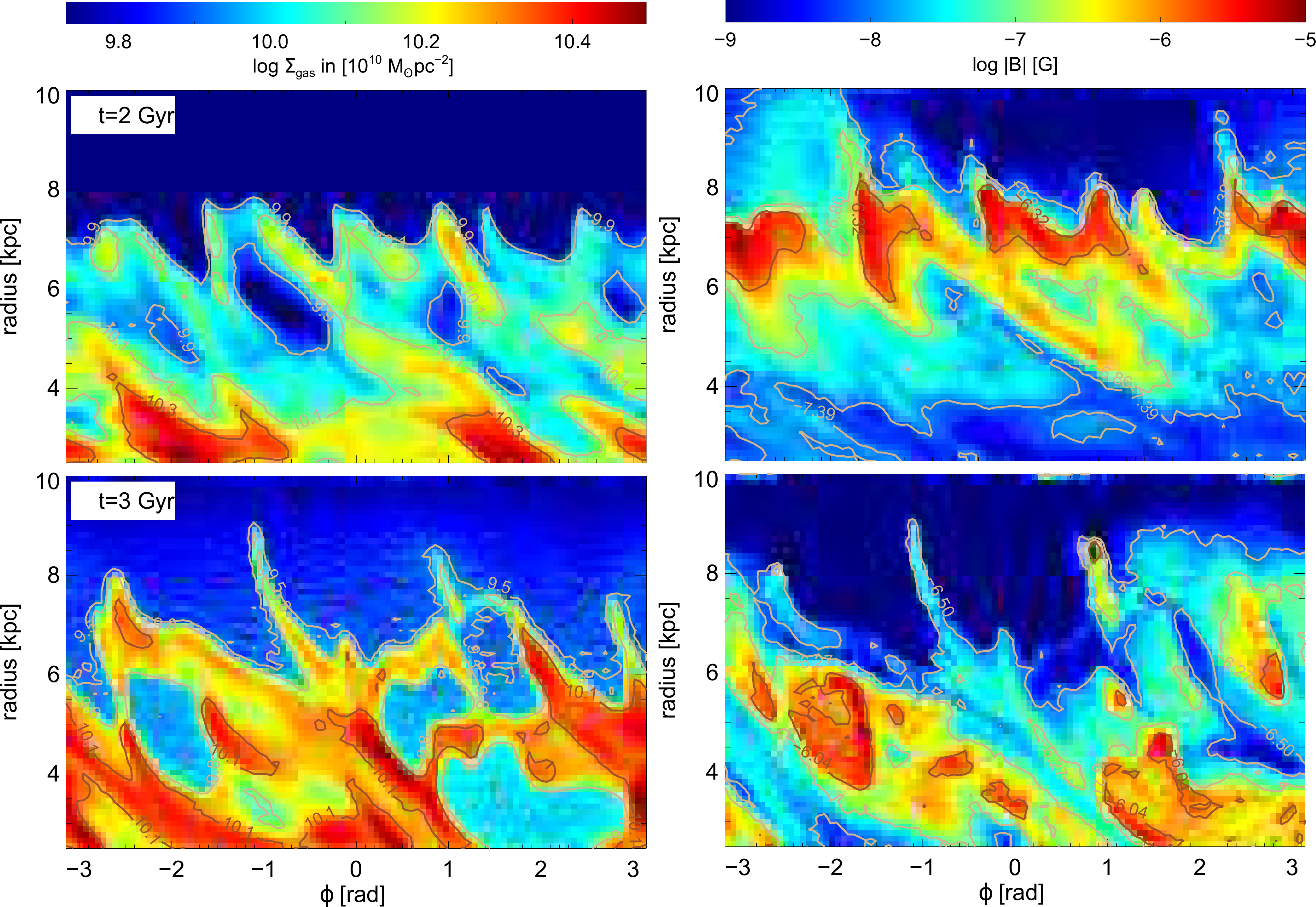}
\caption{Same as figure \ref{fig:rphi1}, but for the simulation \textit{MW-primB}.}
\label{fig:rphi2}
\end{figure*}

The general morphological properties of the magnetic field
have been discussed in section
\ref{sec:morphology}. Here, we present a more detailed study on the galactic magnetic field in our simulations. For this we show a more detailed comparison between the structure of the gas density and the magnetic field strength in polar coordinates 
for the \textit{MW-snB} and \textit{MW-primB} runs in figures \ref{fig:rphi1} and \ref{fig:rphi2}, respectively.
The left-hand side panels show the gas density and the right-hand side panels show the magnetic field strength, while the panels on the top give the results after $t=2$ Gyr and the panels on the bottom give the results after $t=3$ Gyr.
Plotting these quantities in polar coordinates allows us to directly compare our results to observations of spiral galaxies, e.g. by \citet{Bittner2017}.
These plots have been obtained by using a two dimensional grid of the size $100\times66$, with a pixel corresponding to a specific $r$ and $\varphi$. For each pixel we calculate the gas density and the magnetic field using the triangular shaped cloud (TSC) method for calculating densities on a regular grid. The plots then show a slice for a fixed $r$ for one circulation over the whole galaxy. This allows us to
determine the positions of the more prominent areas in the structure of a spiral galaxy.
The panel on the top left of figure \ref{fig:rphi1} nicely shows four density peaks,
corresponding to at least four spiral arms in the \textit{MW-snB} simulation, as expected for a Milky Way analogue. Comparing our results for the gas density to the magnetic field strength, we find that they 
are not strongly correlated. Moreover, early in the simulation, we observe
that the magnetic field is stronger in the outer parts of the galactic disc, compared to the centre. This behaviour changes at later times (bottom panels),
when the magnetic field is stronger in the centre and shows a spiral structure. 
The magnetic field strength is highest between the spiral arms of the gas disc, which is most obvious at late times, when the gas of the disc is already depleted due to star formation and outflows. While \citet{rbeck13} have noted this effect before, they have not found this behaviour in gas rich galaxies. Although, we can give an explanation for the connection between the gas density and the magnetic field strength we note
that our implementation of MHD follows mainly ideal MHD with a small magnetic diffusion term. In the limit of small magnetic fields this leads to a direct correlation between high densities and high magnetic fields. Thus, at the start of the simulation,
the strongest magnetic fields are in the high density areas, due to adiabatic collapse.
At later times the non-linear term we include in the induction equation makes it possible to trigger diffusive processes that can transport the magnetic field to different locations.

\begin{figure*}
\makebox[\textwidth]{%
\includegraphics[width=1\textwidth]{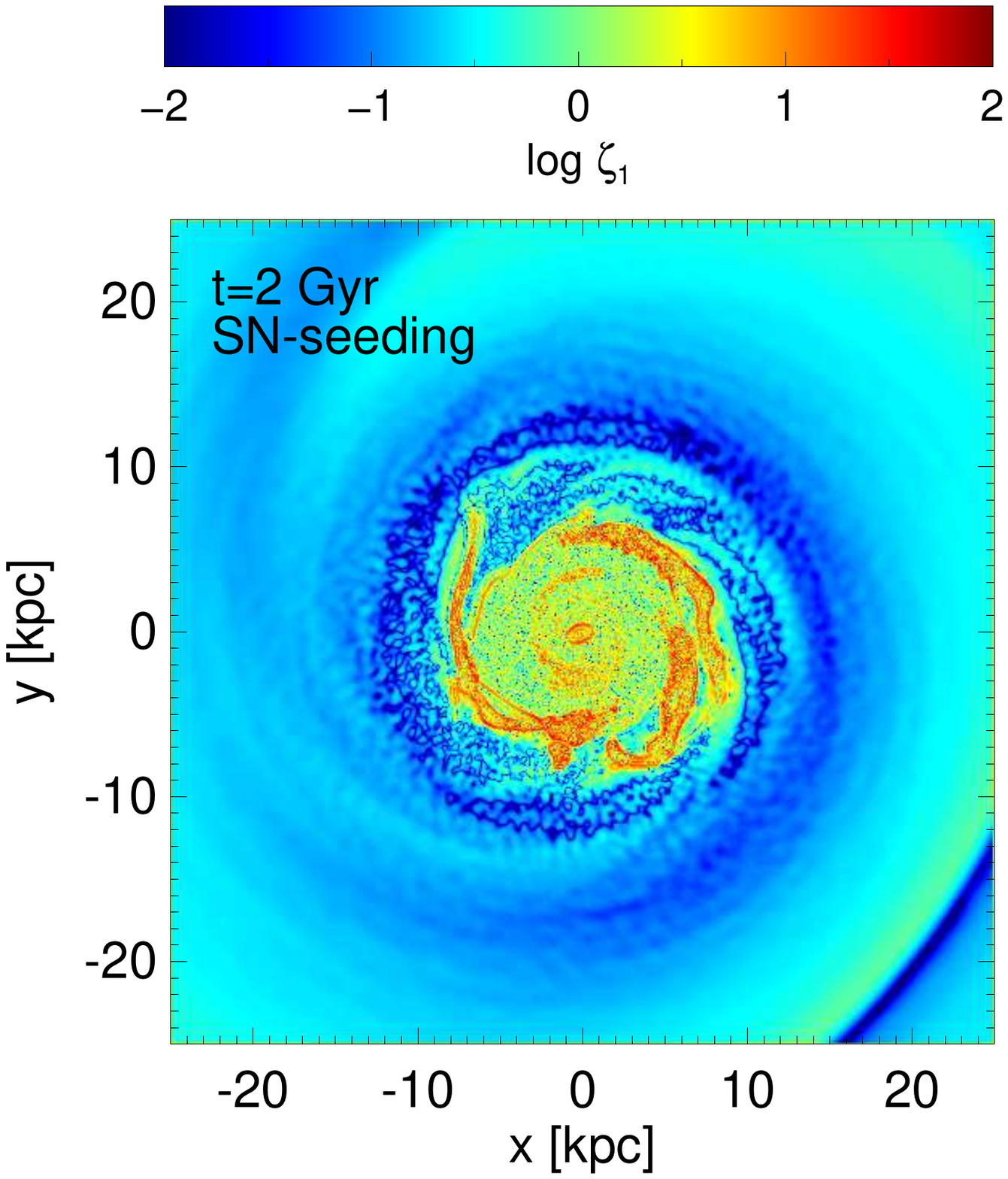}%
\hspace{-9cm}
\includegraphics[width=1\textwidth]{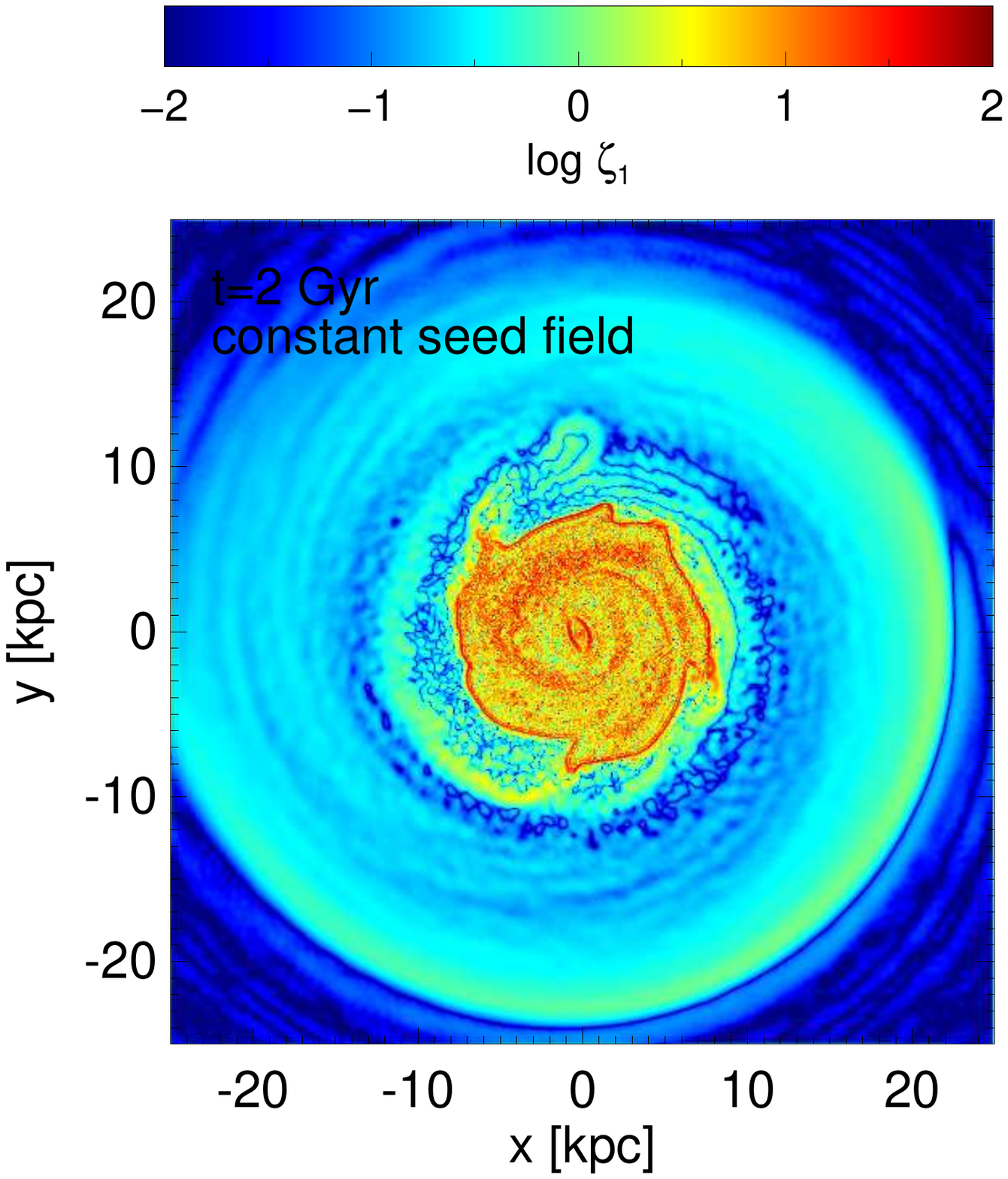}%
}\\[-3.5cm]
\makebox[\textwidth]{%
\includegraphics[width=1\textwidth]{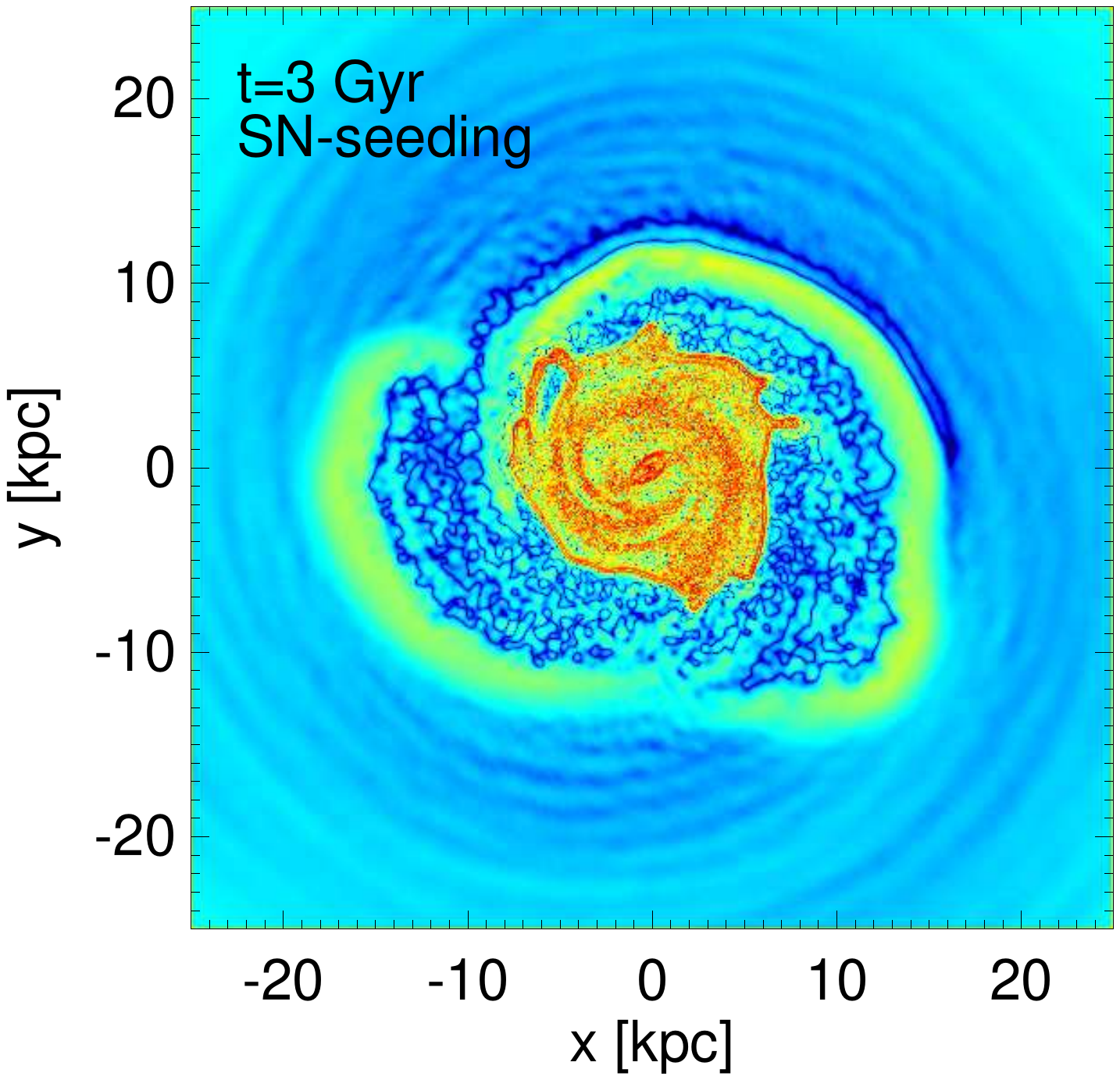}%
\hspace{-9cm}
\includegraphics[width=1\textwidth]{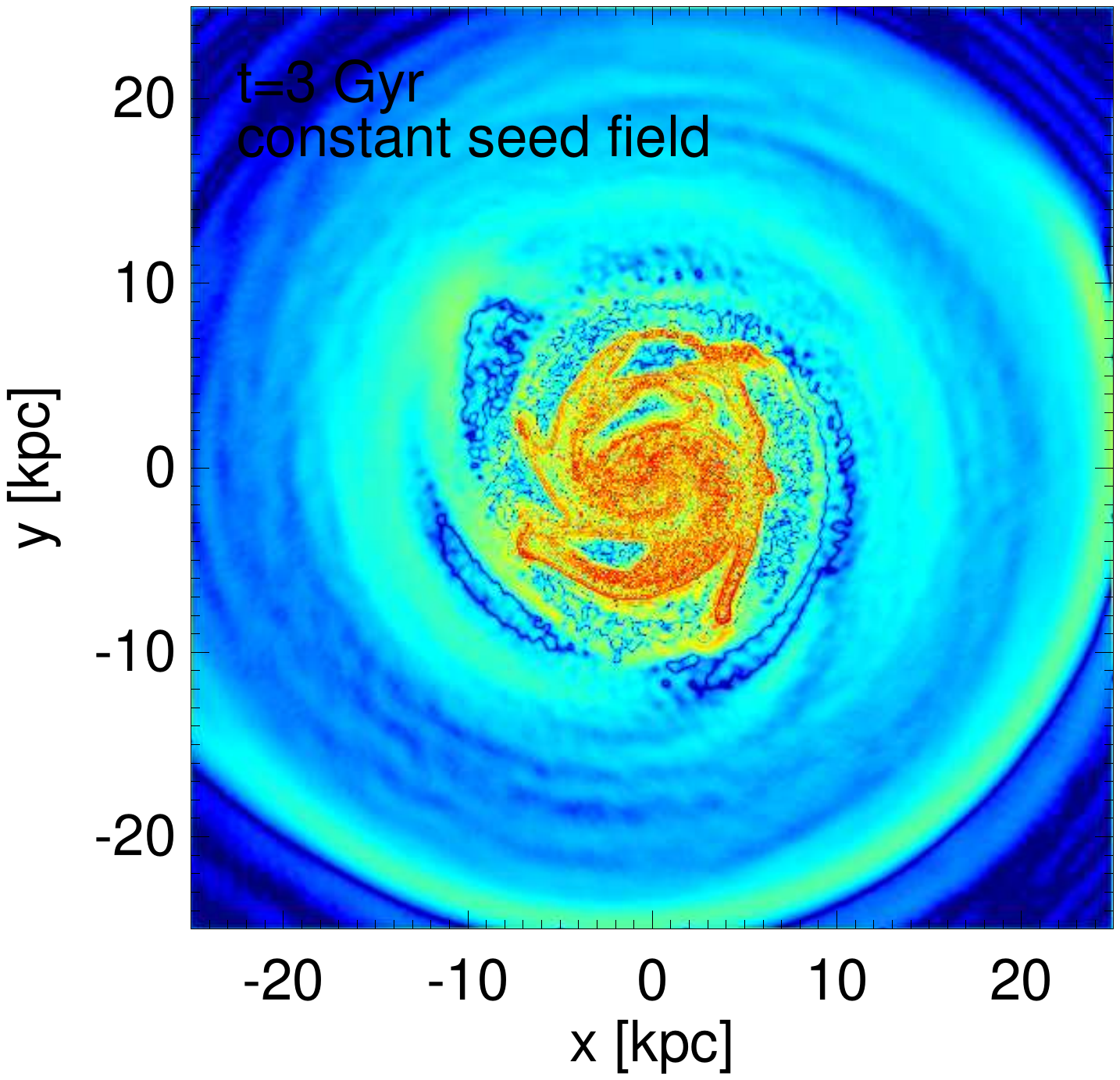}%
}\\[-1.0cm]%
\caption{Structure of the magnetic field for the \textit{MW-snB} run (left-hand-side panels) and the \textit{MW-primB} run (right-hand-side panels) at $t=2$ Gyr (upper panels) and $t=3$ Gyr (bottom panels). The colour bar indicates the normalized relative deviation of the magnetic field from a smoothed magnetic field
(unsharp masking, see eqn. \ref{zeta1}). Red indicates that turbulence is dominating while blue indicates high order in the magnetic field.}
\label{fig:magstruct_method1}
\end{figure*}
 
We now focus on
the detailed structure of the magnetic field and the comparison of the differences in the structure for both presented magnetic field models. We present two ways to evaluate
this.
The first one is based on the evaluation of the quantity $\zeta_{1}$ given by
\begin{align}
    \zeta_{1} = \frac{|\mathbf{B} - \mathbf{B}_{\mathrm{sm}}|}{|\mathbf{B}_{\mathrm{sm}}|}, \label{zeta1}
\end{align}
where $\mathbf{B}$ is the magnetic field and $\mathbf{B}_\mathrm{sm}$ is the magnetic field smoothed by a Gaussian-kernel. As our simulations were performed with an SPH code, $\zeta_{1}$ cannot be directly computed, but the particle properties first need to be transformed to a regular grid. To achieve this, an obvious choice would be the TSC method that was used to obtain figures \ref{fig:rphi1} and \ref{fig:rphi2}. However,
the TSC method is based on a triangular kernel and thus not accurate enough to resolve
the detailed structure of the magnetic field in an SPMHD simulation. Therefore, we
work out an proper SPH-binning for which we use a two dimensional grid with a very high resolution of $1024 \times 1024$ grid points. For each element, we then
calculate the magnetic field with the Wendland C4 Kernel using 200 neighbouring particles. In this way, we use exactly the same configuration to bin the data as for our underlying SPH-formalism.
The data binning has been performed using the code \textsc{sphmapper} presented by \citet{Roettgers2018}. The property $\zeta_{1}$ describes the deviation of the magnetic field in our simulations from the smoothed magnetic field. This method called unsharp masking, is very common in image editing, and allows us to achieve a stronger contrast in the magnetic field, making it is easier to find detailed structure lines.
A high value of $\zeta_{1}$ indicates a large difference between the smoothed magnetic field and the original magnetic field, indicating a highly turbulent magnetic field.
A small value of $\zeta_{1}$ shows little deviation between the smoothed magnetic field and the original one, indicating a region of highly correlated magnetic field lines.

\begin{figure*}
\makebox[\textwidth]{%
\includegraphics[width=1\textwidth]{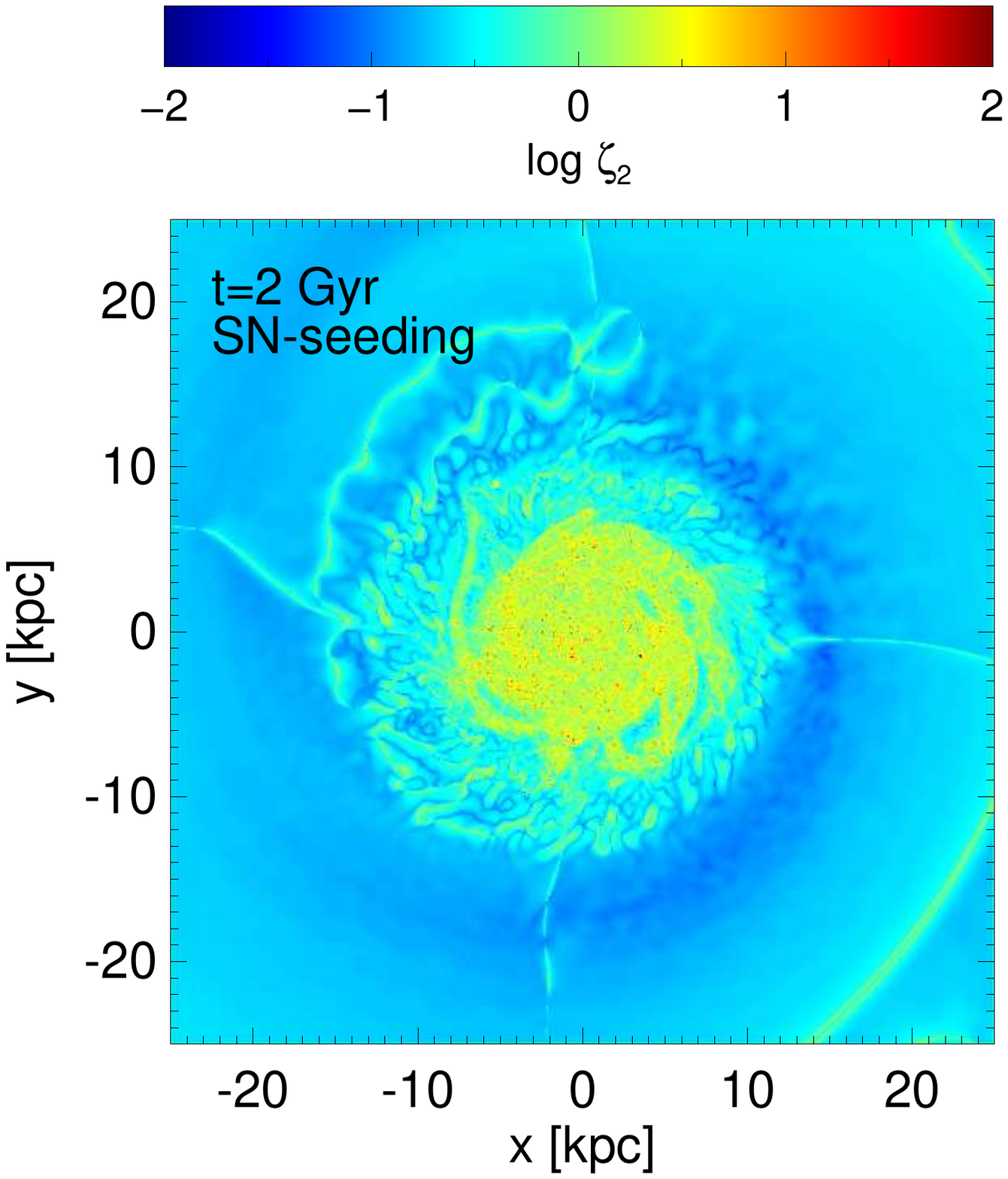}%
\hspace{-9cm}
\includegraphics[width=1\textwidth]{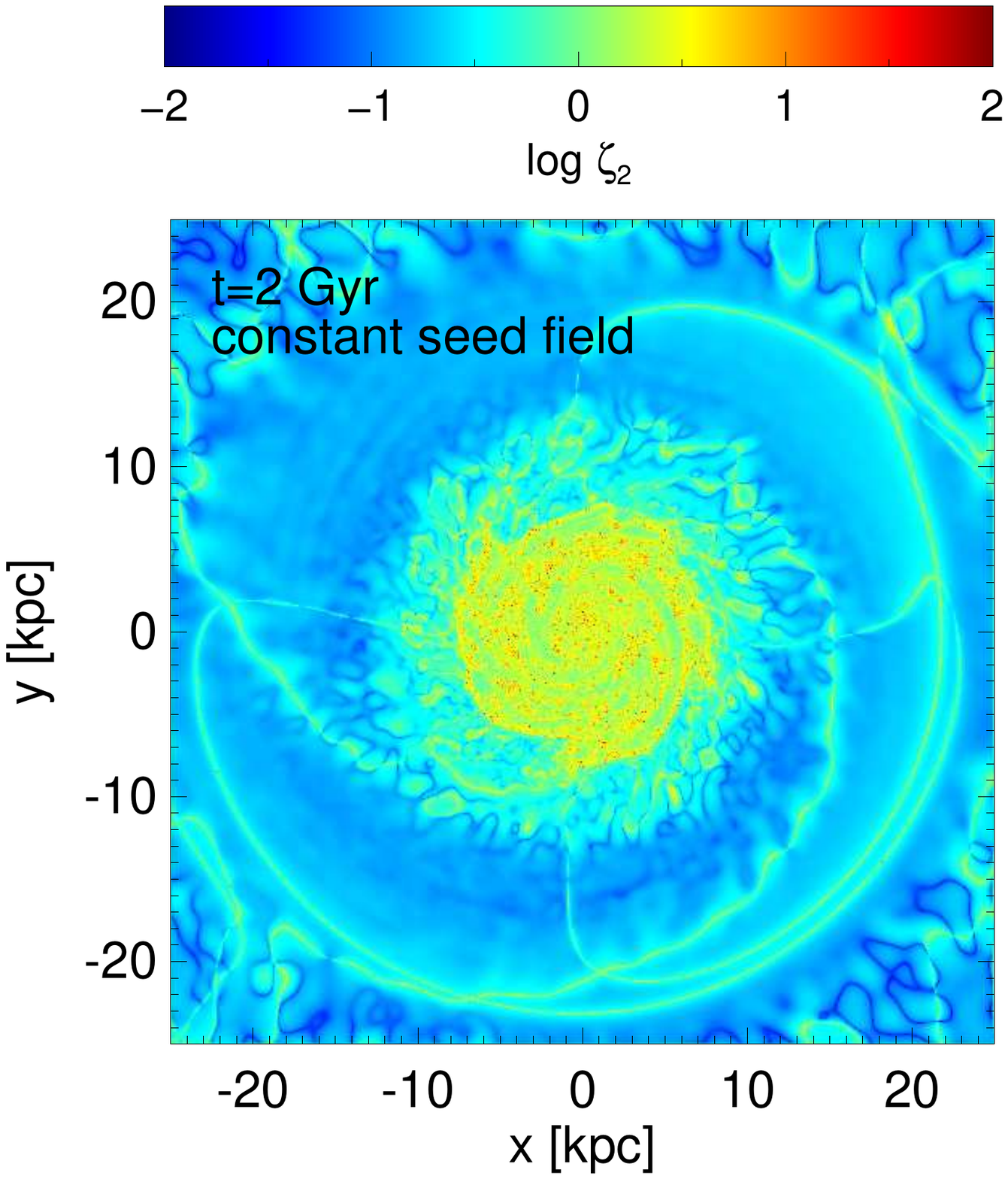}%
}\\[-3.5cm]
\makebox[\textwidth]{%
\includegraphics[width=1\textwidth]{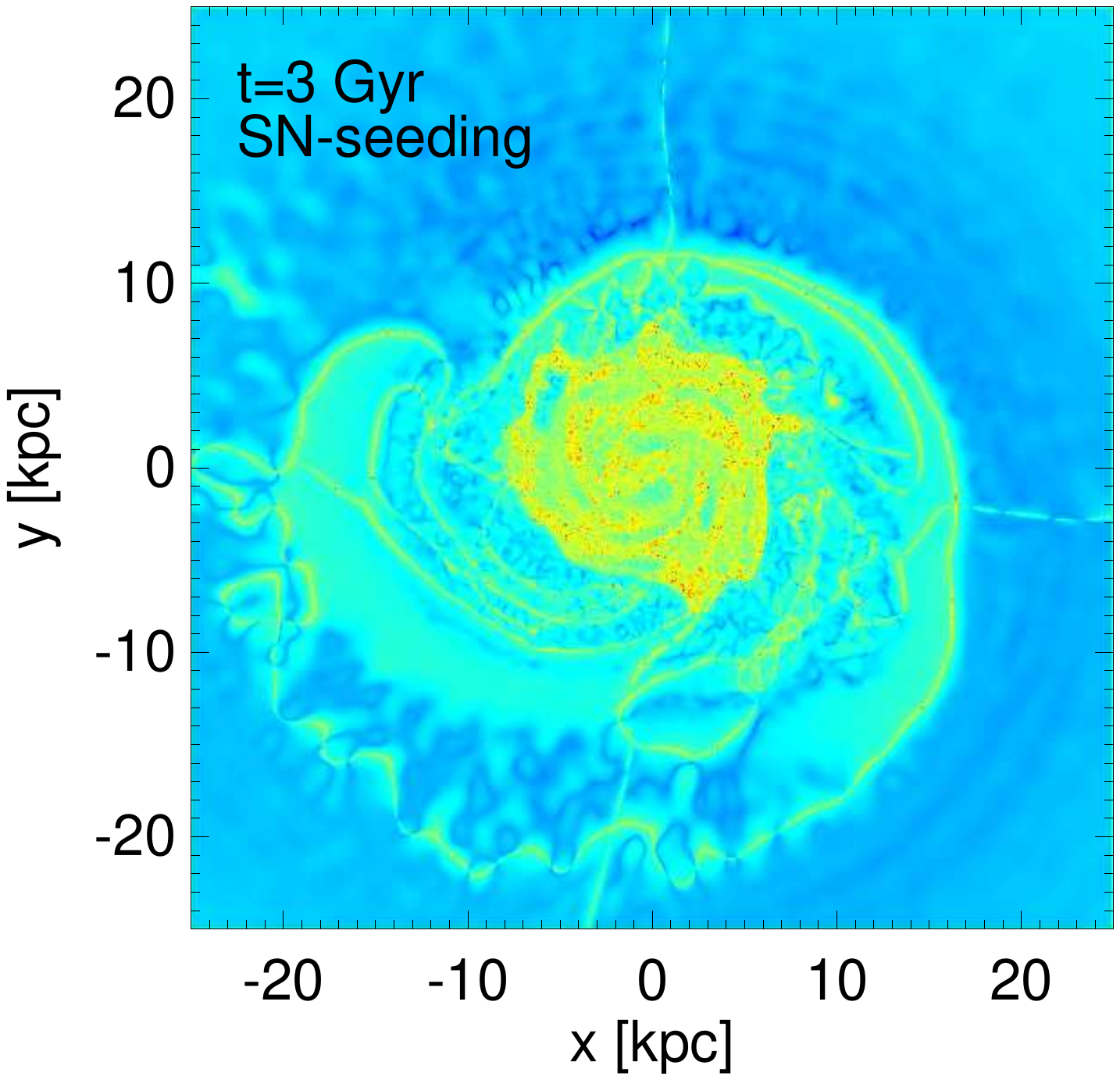}%
\hspace{-9cm}
\includegraphics[width=1\textwidth]{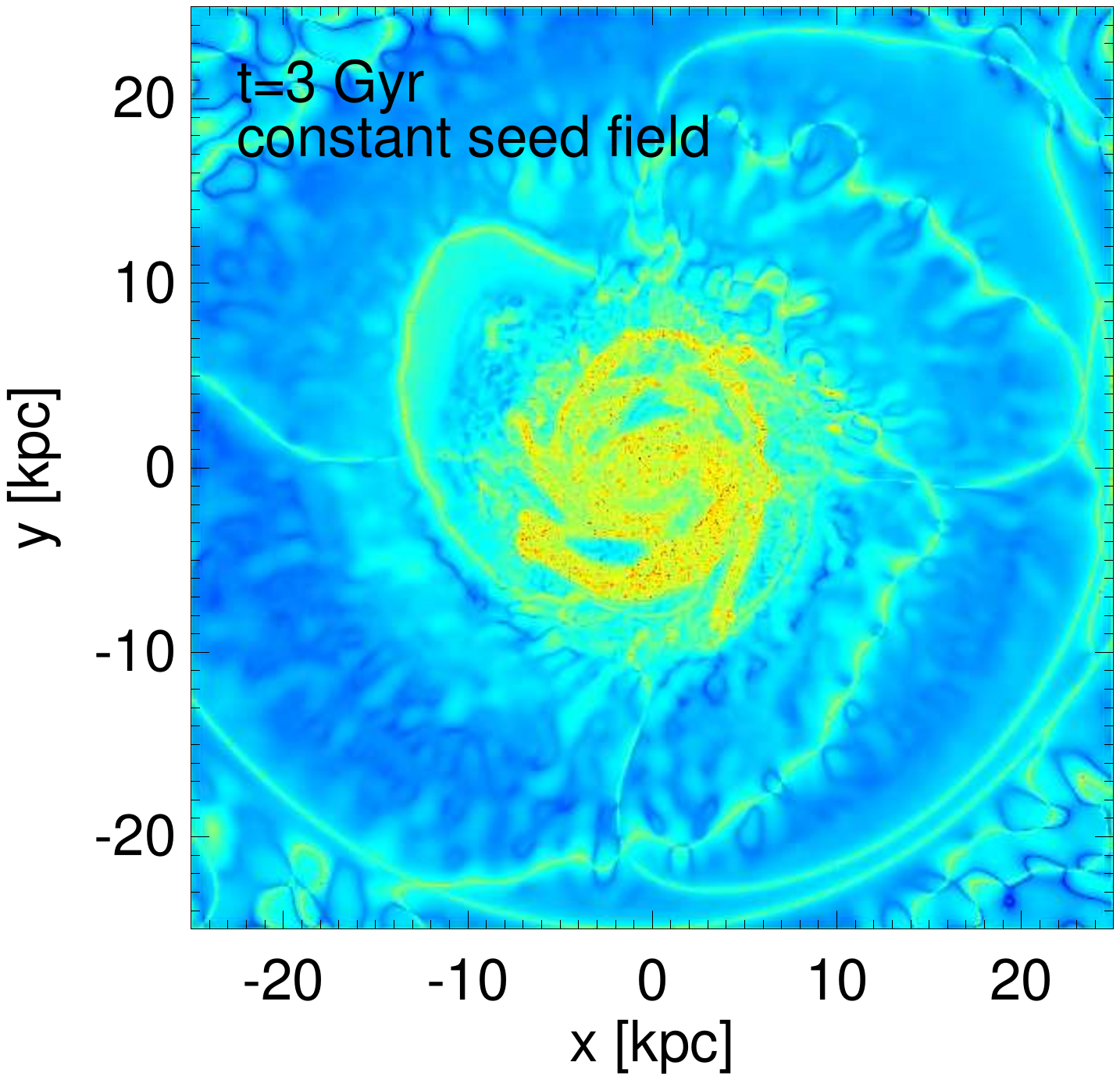}%
}\\[-1.0cm]%
\caption{Same as figure \ref{fig:magstruct_method1}, but each spatial magnetic field direction is normalized to the smoothed magnetic field in that certain direction (see eqn. \ref{zeta2}).}
\label{fig:magstruct_method2}
\end{figure*}

The upper left panel of figure \ref{fig:magstruct_method1} shows $\zeta_{1}$ for the \textit{MW-snB} simulation at $t=2.0$ Gyr, shortly before the outflow sets in.
It demonstrates that there are detailed structures in the magnetic field beyond the spiral arms. Furthermore, the turbulent structures are mostly located in the centre of the galactic plane. This nicely illustrates that the magnetic field is amplified via a small scale dynamo, as we argued in section \ref{sec:amplification}.
On the small scales, i.e. in the galactic centre, the turbulent character of the magnetic field dominates.
The upper right panel of figure \ref{fig:magstruct_method1} shows $\zeta_{1}$ for the \textit{MW-primB} simulation. Also here, we observe the spiral structure of the magnetic field, but the disc is more dominated by turbulence than in the \textit{MW-snB} run.
The bottom panels of figure \ref{fig:magstruct_method1} show $\zeta_{1}$ at $t=3$ Gyr, demonstrating that the structure of the magnetic field is not diminished by the magnetic outflow which sets in around $t=2.3$ Gyr.
The quantity $\zeta_{1}$ evaluates the total magnetic field, ignoring the spatial behaviour of each magnetic field component. We therefore present a second method to evaluate the structure of the magnetic field, and introduce the variable $\zeta_{2}$:
\begin{align}
    \zeta_{2} = \sqrt{\left(\frac{B_{x} - B_{\mathrm{x/sm}}}{B_{\mathrm{x/sm}}}\right)^2 + \left(\frac{B_{y} - B_{\mathrm{y/sm}}}{B_{\mathrm{y/sm}}}\right)^2 + \left(\frac{B_{z} -B_{\mathrm{z/sm}}}{B_{\mathrm{z/sm}}}\right)^2}.\label{zeta2}
\end{align}
Compared to $\zeta_{1}$, it takes into account the structure of the magnetic field in each spatial direction, and prevents an overestimation due to the strong magnetized outflows in z-direction. The resulting maps are presented in figure \ref{fig:magstruct_method2}. While the $\zeta_1$ maps clearly show the underlying magnetic structure of the disc in the x-y-plane, the $\zeta_2$ maps are far smoother, though we can see the structures we already saw in figure \ref{fig:magstruct_method1}. 
Finally, we note that the magnetic field becomes much more turbulent when the magnetic outflow sets in.  

\subsection{Different halo masses \label{sec:halomasses}}
 
So far, we focused on the simulations \textit{MW-noB}, \textit{MW-snB} and \textit{MW-primB}, as these systems were constrained by X-ray observations. Here, we investigate the effects in haloes of lower mass, i.e. 
$M_\mathrm{h}=10^{11}M_{\odot}$ (\textit{MM-noB}, \textit{MM-snB}, \textit{MM-primB}) and $M_\mathrm{h}=10^{10}M_{\odot}$ (\textit{DW-noB}, \textit{DW-snB}, \textit{DW-primB}). While the intermediate mass systems (MM) show a similar behaviour
as the Milky Way-like systems (MW) for the amplification of the magnetic field and the observed morphological features, the point in time where the biconical magnetic tube sets in is is delayed to $t=3.0$ Gyr. This is a consequence of the magnetic field amplification being driven by small scale turbulence induced by feedback, mainly in the galactic centre. In the lower mass galaxies the efficiency of the feedback is lower,
resulting in a slower amplification of the magnetic field. 
Moreover, the lower rotational velocity in a halo with $M_\mathrm{h}=10^{11}M_{\odot}$
delays the amplification process further. 
As a consequence, the magnetic pressure rises at a lower rate compared to the Milky Way-like systems, such that the magnetic pressure needed to expell gas from the centre towards the CGM is reached at a later point in time. The total magnetic field strength rises to similar values as for the Milky Way-like models, but the peak values of the magnetic field is slightly lower, reaching values between $10^{-7}$ to $10^{-5}$ G.
While the magnetic field properties of the MM simulations are similar to the MW simulations, i.e. a small scale turbulent dynamo drives the amplification of the magnetic field resulting in an outflow of gas, we notice a considerable difference in the evolution of the SFR.
For the system with the lowest mass of $M_\mathrm{h}=10^{10} M_{\odot}$ (DW), we only find minor changes for the simulations with magnetic fields compared to the reference simulation without magnetic fields. For this halo mass scale the magnetic field is dynamically unimportant, as the amplification of the magnetic field  is very slow. 
Since the SFR in these systems is very low, the effects of the SN feedback are minor, resulting in no significant small scale turbulence that could amplify the magnetic field in the centre. 
Moreover, the amplification of the magnetic field can not be supported via the $\alpha$-$\omega$-dynamo either, because the gas orbits with a peak velocity of $50$ km/s. 
Lastly, the magnetic field can also not be amplified by adiabatic compression of the gas because the potential wells are too shallow to trigger the formation of high density regions which would result in an amplification of the magnetic field.
Consequently, there are no outflows for the DW simulations. We conclude that outflows are only present when the magnetic field becomes dynamically important. This is the case for the MW and MM haloes, because the amplification of the magnetic field is strong enough to reach a magnetic pressure that is higher than the thermal pressure.

\subsection{Divergence Cleaning \label{sec:divergence_cleaning}}

\begin{figure*}
        \centering
        \includegraphics[scale=0.5]{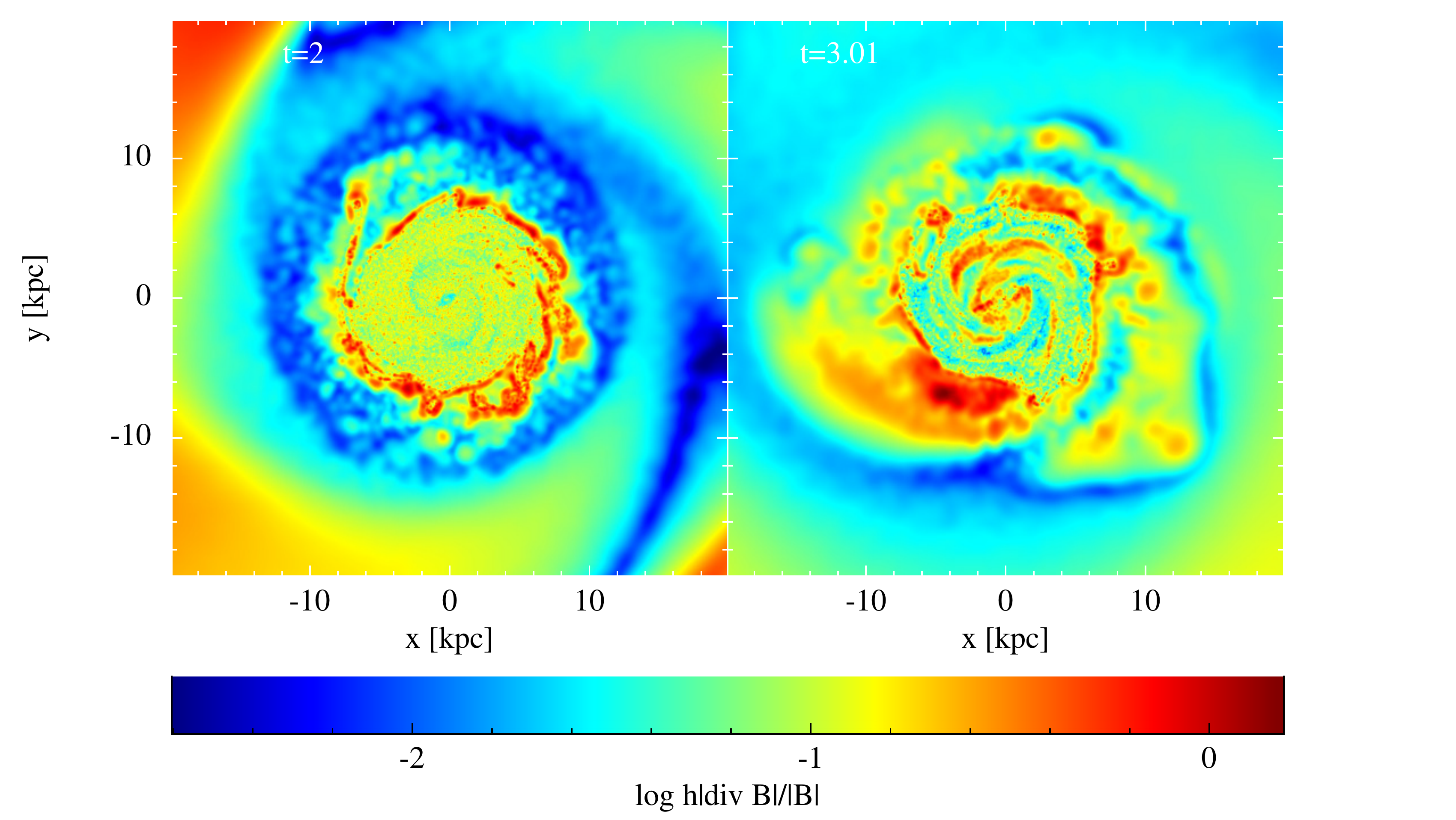}
        \caption{Relative divergence errors for the simulation \textit{MW-snB} at $t=2$ Gyr (left-hand-side panel) and $t=3$ Gyr (right-hand-side panel). Typical values are between 1 and 10 per cent.
        \label{fig:divCleaning1}}
        \includegraphics[scale=0.5]{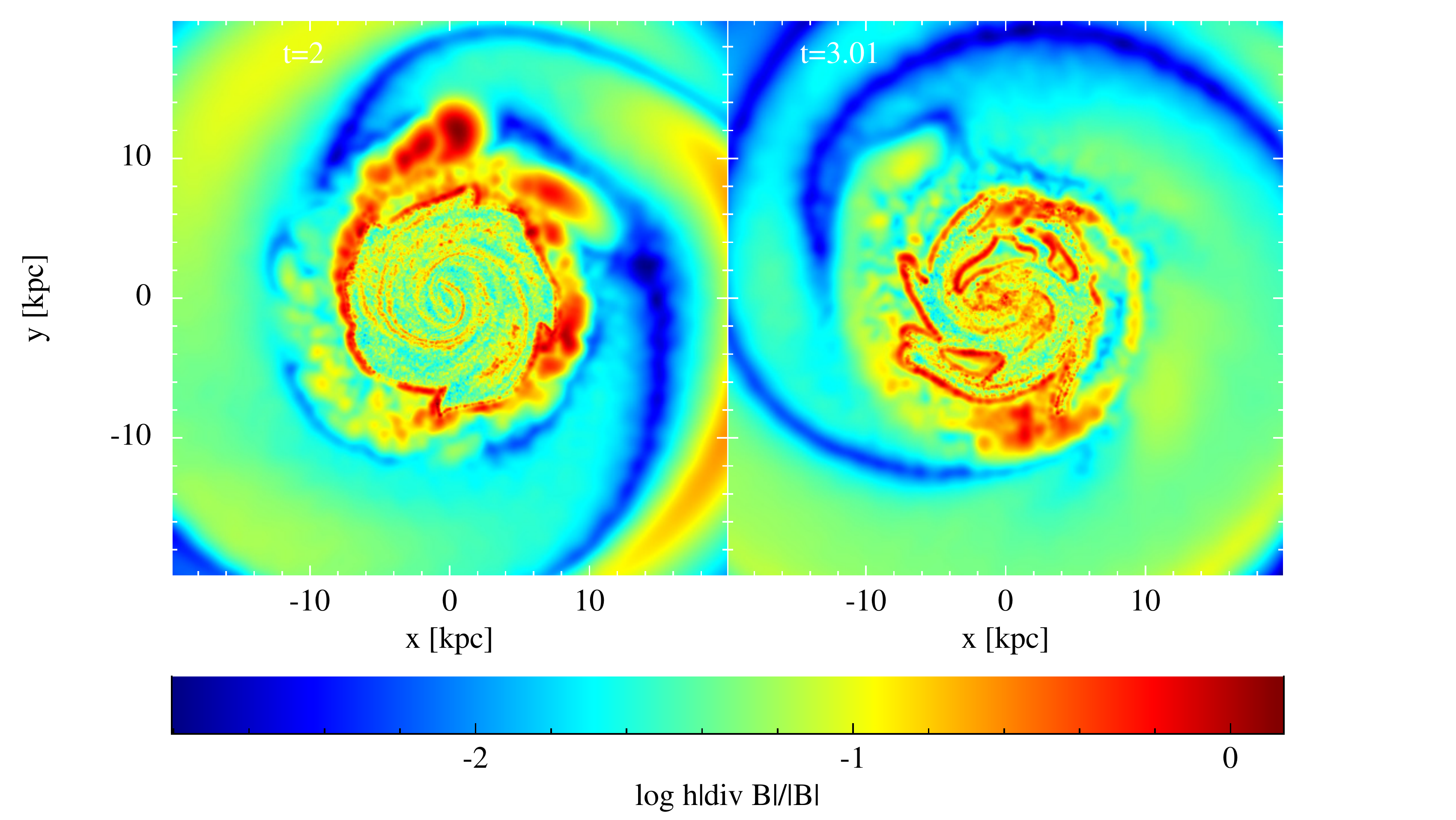}
        \caption{Same as figure \ref{fig:divCleaning1} but for the simulation \textit{MW-primB}. \label{fig:divCleaning2}}
\end{figure*}

Every numerical simulation that includes magnetic fields has to deal with the $\nabla \cdot \mathbf{B} = 0$ constraint, we need to prove that our simulations are not contaminated by magnetic monopoles. For this, we present the relative magnetic divergence $h \cdot \mathbf{\nabla} \mathbf{B} /\vert\mathbf{B}\vert$, 
where $h$ is the smoothing length, for the \textit{MW-snB} run in figure \ref{fig:divCleaning1}, and for \textit{MW-primB} in figure \ref{fig:divCleaning2}. 
The panels on the left-hand-side and right-hand-side show the result before ($t=2.0$ Gyr), and after ($t=3.0$ Gyr)
the magnetic outflow sets in.
The more turbulent structure after the outflow sets in
leads to an increase of the divergence of the magnetic field due to the sharper gradients. The maximum value of the relative divergence error of $\sim1$ is reached where turbulence is dominating, such as in the galactic centre, and at the edges of the spiral arms. This is comparable to the error in other simulations \citep[e.g.][]{Pakmor2013}\footnote[1]{\textbf{We note that the magnetic energy density is about a factor of $10^{-2}$ below the kinetic energy density within one resolution element of the simulation in its very centre.}}. However, the typical value of $h \cdot \mathbf{\nabla} \mathbf{B} /\vert\mathbf{B}\vert$ is below 0.1, which is very good for a SPMHD-Code. We note that there are methods in grid codes such as \textsc{RAMSES}, which can reduce the relative divergence error to machine precision. Although, these cleaning methods perform very well in test studies they have the disadvantage of being computationally very expensive. However, \citet{Pakmor2013} used a cleaning scheme \citep{Powell1999} similar to ours and find that it is sufficient on a moving mesh. Compared to a regular grid, it is significantly more complicated to control the divergence error on amoving mesh because of the irregularity of the Voronoi-grid, which is similar for a SPH particle distribution. Although our SPMHD formulation includes higher order hyperbolic cleaning schemes \citep{Dedner2002, dolag09},
the findings of \citet{Pakmor2013} show that it is sufficient to use a lower order cleaning scheme for the type of systems we are simulating. To save computational power we follow this approach and use a lower order cleaning scheme.
We can further observe that the higher values of the relative divergence error appear
around the spiral arms, as this is where the magnetic field has strong gradients.
This behaviour improves with higher resolution, because the gradients become better resolved. Lastly, we note that the divergence is smaller in the case of the supernova-seeding, since magnetic dipoles are inserted into the ISM. By construction, this leads to a lower divergence, as the dipole structure is forced to appear with the supernova explosions, resulting in a smoother distribution of the magnetic field.

\section{Conclusions} \label{sec:conclusions}

We present a modified model for isolated disc galaxies including a realistic CGM. Using this model we simulate a set of galaxies with different halo masses ranging from $10^{10} M_{\odot}$ over $10^{11} M_{\odot}$ to $10^{12} M_{\odot}$ and study
the general properties (like the morphological structure and the SFR) of these systems.
We focus on the Milky Way-like system, and present a detailed study of the morphological structure of the gas density, as well as the magnetic field. We observe a mean magnetic field strength of a few $\mu$G in the galactic disc, which is in good agreement with observations. In the galactic centre we find higher field strengths up to a few $100 \mu$G. In the spiral arms the magnetic field strength is about an order of magnitude lower compared to the galactic centre. We find that the structure of the magnetic field strength does not follow exactly the spiral arms in the gas density, but is strongest
between two neighbouring spiral arms. This result differs from those
reported by other groups \citep{Pakmor2013, Butsky2017}. The reason for that is that we include a magnetic diffusion term in our simulations, which makes it possible to follow the magnetic field evolution in the non-linear regime. Furthermore, this effect is in agreement with many observations (see \citet{Beck2015} for and references therein).
The amplification of the magnetic field in our simulations is mainly driven by small scale turbulence. We show clear evidence for this in the magnetic power spectra
(figure \ref{fig:power}), in agreement with simulations by other groups \citep{Pakmor2013, Rieder2016, Butsky2017}. 

Moreover, we find further evidence for a small scale dynamo by examining the curvature of the magnetic field lines, which can be used to distinguish between amplification by adiabatic compression and by a small scale dynamo (figure \ref{fig:curvature}). Our simulations indicate that at later times the slope of the magnetic power spectra turns around. This shows that the galaxies are entering a new regime that is dominated by strong magnetic fields instead of small scale turbulence. Thus, the dominating amplification process in this regime is either driven by the $\alpha-\omega$-dynamo or completely saturated and thus switched off.
In the simulations with $M_\mathrm{h}=10^{12} M_{\odot}$ and $M_\mathrm{h}=10^{11} M_{\odot}$, we find galactic outflows that are driven by the magnetic field. In this regime the magnetic pressure is several orders of magnitude higher than the thermal pressure. In our simulations with $M_\mathrm{h}=10^{10} M_{\odot}$ this is not the case,  so that we do not observe a dominating magnetic outflow in haloes below $M_\mathrm{h}=10^{11} M_{\odot}$. A more detailed study of the interaction between galactic disc and CGM shows that a certain amount of magnetic energy is released in the outer regions of the CGM having its origin in the centre of the galactic disc. Studying the turbulence in the magnetic field, we find that the highly magnetized outflows are mainly driven by the turbulent magnetic field in the centre of the galactic disc. The structural analysis of the magnetic field indicates that it follows a complex structure besides the obvious spiral patterns and does not necessarily follow the spiral structure of the gas density because of magnetic diffusion. Finally, we summarize the three most important findings of this study.

\begin{itemize}
	\item Amplification of the magnetic field strength is driven by small scale turbulence until the magnetic field in the disc is strong enough so that the dynamo saturates. We provide evidence for this \citet{Kazantsev1968} dynamo in the magnetic power spectra as well as the anti-correlation of the magnetic field strength and the curvature of the magnetic field lines for a small scale dynamo \citep[e.g.][]{Schekochihin2004, Vazza2018}.\\
	
	\item Galaxies in which the magnetic pressure dominates the thermal pressure show magnetic-driven outflows that can lead to a significant mass loss of the baryonic disc. The outflows appear as low density bubbles reaching a several $100$ km/s before they mix with the CGM and fall back to the disc.\\
	
	\item Diffusive terms in the inducution equation can lead to an anti-correlation between the spiral structure of the gas disc and the spiral structure within the magnetic field strength that can be seen in observations.
\end{itemize} 

Future work will need to focus on detailed resolution studies to determine the spatial and the mass resolution that is needed to actually resolve the small scale turbulent dynamo, which may be crucial in the framework of cosmological zoom-in simulations of Milky Way-like galaxies. Furthermore, none of the current models for star formation in hydrodynamical simulations includes the pressure given by the magnetic field directly.
Only indirect effects on the SFR can be captured by the current simulations. In future studies the magnetic pressure may be directly included by using pressure-based star formation models.       


\section*{Acknowledgments}

We thank Eirini Batziou, Andreas Burkert, Julien Devriendt, Thorsten Naab, Ruediger Pakmor, Rhea Silvia Remus, Felix Schulze, Romain Teyssier and Simon White for useful discussions and their insights on magnetic fields and galaxies. We thank Franco Vazza for his insights on the magnetic curvature. The authors gratefully acknowledge the computing time granted by the John von Neumann Institute for Computing (NIC) provided on the supercomputer JURECA at J\"ulich Supercomputing Centre (JSC) under the project number hmz07 and the computing time provided by the Leibniz Rechenzentrum (LRZ) of the Bayrische Akademie der Wissenschaften on SuperMuc in Garching with the porject number pe86re. UPS and BPM are funded by the Deutsche Forschungsgemeinschaft (DFG, German Research Foundation) with the project number MO 2979/1-1. KD acknowledges support by the DFG Cluster of Excellence 'Origin and Structure of the Universe'


\bibliographystyle{mnras}
\bibliography{paper}

\bsp
\label{lastpage}
\end{document}